\documentclass[journal abbreviation, manuscript]{copernicus}
\usepackage{enumitem}
\usepackage{amsmath}
\usepackage{graphicx}
\usepackage{verbatim}
\usepackage{hyperref}
\usepackage{tabularx}
\usepackage{booktabs}
\begin{document}\nolinenumbers

\title{Reconstructing Three-decade Global Fine-Grained Nighttime Light Observations by a New Super-Resolution Framework}

\Author[1,2]{Jinyu}{Guo}
\Author[1,2]{Feng}{Zhang}
\Author[2,3]{Hang}{Zhao}
\Author[2,4]{Baoxiang}{Pan}
\Author[5]{Linlu}{Mei}

\affil[1]{Department of Atmospheric and Oceanic Sciences \& Institute of Atmospheric Sciences, Fudan University, Shanghai 200438, China}
\affil[2]{Shanghai Qizhi Institute, Shanghai 200030, China}
\affil[3]{Institute for Interdisciplinary Information Sciences, Tsinghua University, Beijing 100084, China}
\affil[4]{Institute of Atmosphere Physic, Chinese Academy of Sciences, Beijing 100029, China}
\affil[5]{Institute of Environmental Physics, University of Bremen, Bremen 28359, Germany}
\correspondence{Feng Zhang (fengzhang@fudan.edu.cn)}
\runningtitle{Reconstructing Three-decade Global Fine-Grained NTL Observations by a New SR Framework}
\runningauthor{Jinyu Guo et al.}

\firstpage{1}

\maketitle

\begin{abstract}
Satellite-collected nighttime light provides a unique perspective on human activities, including urbanization, population growth, and epidemics. Yet, long-term and fine-grained nighttime light observations are lacking, leaving the analysis and applications of decades of light changes in urban facilities undeveloped. To fill this gap, we developed an innovative framework and used it to design a new super-resolution model that reconstructs low-resolution nighttime light data into high resolution. The validation of one billion data points shows that the correlation coefficient of our model at the global scale reaches 0.873, which is significantly higher than that of other existing models (maximum = 0.713). Our model also outperforms existing models at the national and urban scales. Furthermore, through an inspection of airports and roads, only our model's image details can reveal the historical development of these facilities. We provide the long-term and fine-grained nighttime light observations to promote research on human activities. The dataset is available at \url{https://doi.org/10.5281/zenodo.7859205}.
\end{abstract}

\introduction
    Satellite-collected nighttime light (NTL) data can depict the spatial distribution and strength of artificial light sources on the earth's surface, providing a distinct perspective for studying various facets of human activities \citep{TAN2022112834,Liu9665790,Hu9366289}.Defense Meteorological Satellite Program’s Operational Linescan Systems (DMSP-OLS) are the sensors installed on a family of satellites \citep{ma2020constructing,Zhao8897687,ZHENG201936}. These sensors produced the longest-term NTL archives from 1992 to the present and have become the main source of NTL data \citep{BENNETT2017176}.  DMSP-OLS is widely used in the analysis of carbon emissions and light pollution \citep{SHI2016523,MohsinJamilButt}, estimation of gross domestic product (GDP) and population \citep{WU2013111,LI20181248}, observation of conflicts and disasters \citep{Frank,Xue}, and mapping of built-up areas and impervious surfaces \citep{ZHOU2014173,ZHUO201864}, etc. Nevertheless, despite its usefulness, DMSP-OLS data has a limitation of coarse granularity which hinders its applicability for in-depth analysis of urban facilities \citep{Cao9507080}.

One reason for the coarse granularity is its low spatial resolution of 1 km \citep{elvidge2013viirs}. More importantly, it has problems of overglow effect and saturation \citep{rs11171971}. The overglow effect refers to the significant impact of bright pixels on surrounding areas \citep{Sun7934422}. This effect not only results in images appearing too smooth, obscuring details within cities, but also causes some areas without light sources, such as the sea surface, to be illuminated by nearby cities \citep{ZheyanShen}.  The saturation problem means that the quantization capacity of DMSP-OLS is only 8 bits, so the brightness of urban centers is stable at 63 and has not changed \citep{ZihaoZheng,9316779,9366289}. In response to the overglow effect, thresholding methods and classification algorithms were used to identify and eliminate bright pixels without light sources \citep{Henderson,CAO20092205,ZHOU2014173}. Additionally, some studies assumed that the overglow effect was governed by the point spread function and thus used deconvolution filters to sharpen the image \citep{ABRAHAMS2018242,ZHENG2020111707}. Another assumption was that the pixel-level overglow effect was linearly cumulative, so a self-adjusting model was used to correct the image \citep{CAO2019401}. In dealing with the saturation problem, multi-source data represented by vegetation indices is considered correlated with NTL, and many studies fused DMSP-OLS data with the multi-source data to increase urban interior details \citep{rs13061171,rs71215863,rs12233988,POK2017104}. All of these efforts tried to cope with the problem of coarse granularity. However, most of them were based on strong assumptions, which were not necessarily consistent with facts. Therefore, their results are inadequate, and it is necessary to develop new methods to effectively solve the coarse granularity issue.

Another widely used NTL sensor is the Suomi National Polar-orbiting Partnership Satellite’s Visible Infrared Imaging Radiometer Suite (NPP-VIIRS) \citep{XiLidoi:10,Xi_Shao}. It spans from April 2012 to the present and has a spatial resolution of approximately 500 m \citep{ZHENG2021129488}, as well as onboard calibration \citep{8790968}. Compared with the previous generation of sensors, its overglow effect is much less significant \citep{8897687}. Additionally, its quantization capacity is 14 bits, so there is no saturation problem \citep{ZHENG201936}. Overall, NPP-VIIRS is more advanced, and its data has the advantage of being continuous and fine-grained. However, the short time range makes it difficult to be used for the long time-series analysis and applications.

In recent years, some studies have attempted to reconstruct DMSP-OLS data into NPP-VIIRS data to advantageously combine the long-term temporality of DMSP-OLS with the fine-grained nature of NPP-VIIRS. One such study fused the Moderate Resolution Imaging Spectroradiometer (MODIS) Normalized Difference Vegetation Index (NDVI), and DMSP-OLS data to get the vegetation-adjusted NTL urban index (VANUI), which consists of more details. Then a power function was used to establish the regression relationship between UANUI and NPP-VIIRS \citep{YingTu}. Subsequently, the Random Forest and Multilayer Perceptron models were adopted in the reconstruction process. The input variables included DMSP-OLS data, Digital Elevation Model (DEM), and road map \citep{SumanaSahoo}. The models they employed adhered to a point-to-point paradigm, where the spatial relationship among pixels was disregarded. As a result, their application is confined to a city-scale level, and their accuracy warrants further improvement. Super resolution, a technique that reconstructs blurred images into clear images, has made great progress in the last decade with the advent of deep learning \citep{8723565,CHEN2022124}. AutoEncoder, a deep learning model, was modified and used to reconstruct DMSP-OLS data to NPP-VIIRS data \citep{essd-13-889-2021}. This is an image-to-image model that utilizes the spatial relationship among pixels. It is the first model to create global VIIRS-like images. Nevertheless, owing to the reliance on the MODIS Enhanced Vegetation Index (EVI), DMSP-OLS archives before 2000 were not exploited. Moreover, the model’s generalization ability degrades in untrained years, as shown in the Result Section, which causes unclear output images in earlier years. In general, all the current models establish a direct mapping from DMSP-OLS data and associated variables to NPP-VIIRS data of the same year. Despite some advancements in this field, there remains a dearth of long-term and fine-grained observations of NTL.

This study proposes a novel approach to effectively reconstruct global fine-grained NTL observations covering a period of three decades. We show that both statistical and visual performances of our methodology are significantly better than those of existing methods through validation at the urban, national, and global scales. Moreover, our model reveals the long-term historical changes in some certain facilities that cannot be detected by existing models. The main contributions of this study are as follows: 
\begin{enumerate}
    \item We derived a new super-resolution framework and designed a new deep learning model, DeepNTL, to reconstruct DMSP-OLS data into NPP-VIIRS data;
    \item We evaluated our model and found it is superior to existing models at different scales, and can reveal the decades of light changes in urban facilities;
    \item To our knowledge, it is the first time to achieve and release a global fine-grained NTL dataset from 1992 to the present with the highest accuracy achieved so far.
\end{enumerate}

The rest of this paper is organized as follows: Section 2 introduces materials and the continuity correction for DMSP-OLS; Section 3 describes the new super-resolution model; Section 4 illustrates implementation and experiments; Section 5 evaluates our result comprehensively; Section 6 presents the summary and conclusions.

\section{Materials}
    \begin{figure}[!htbp]
    \centering
    	\includegraphics[width=1.0\textwidth]{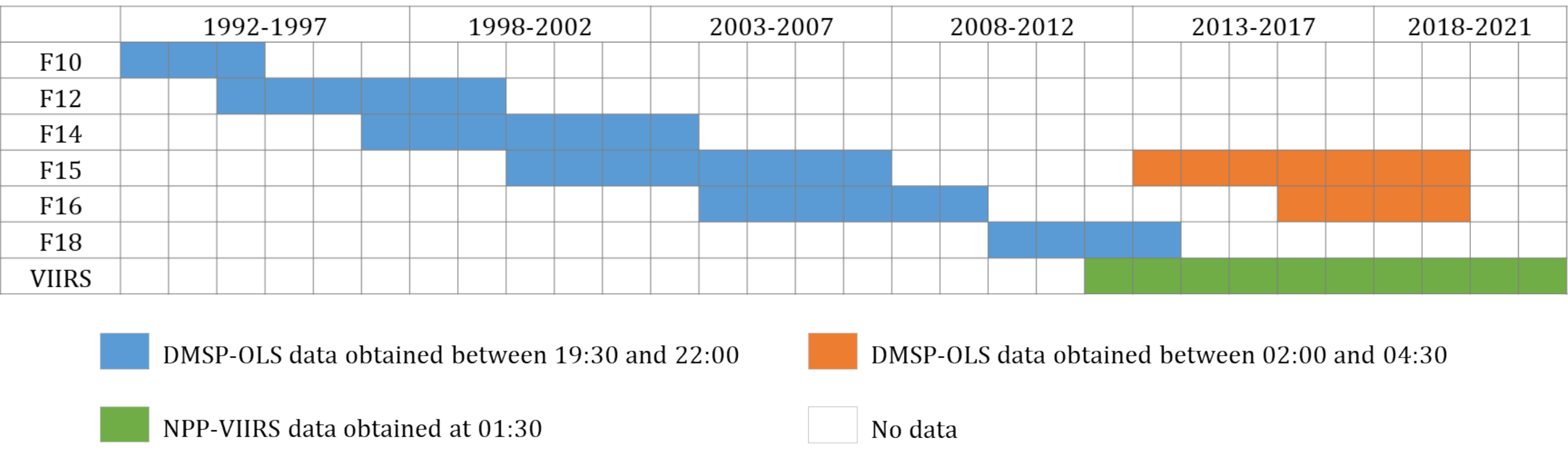}
    	\caption{DMSP-OLS and NPP-VIIRS data. F10 to F18 are the satellite codes of DMSP-OLS. VIIRS is short for NPP-VIIRS.}\label{fig01}
\end{figure}

There are 6 satellites carrying DMSP-OLS sensors, and they are coded as F10, F12, F14, F15, F16, and F18 \citep{rs13245026}. From 1992 to 2013, their overpass times were between 19:30 and 22:00, as represented in blue in Figure \ref{fig01}. After 2013, the orbits of F15 and F16 drifted and their overpass times changed to between 02:00 and 04:30 after midnight, as shown in orange \citep{rs13245004}. The Earth Observation Group of the Colorado School of Mines released Version 4 DMSP-OLS Nighttime Lights Time Series\footnote{https://eogdata.mines.edu/products/dmsp/}. There are three types of products in the dataset: \textit{cf\_cvg}, \textit{avg\_vis}, and \textit{stable\_lights.avg\_vis}. The \textit{cf\_cvg} product records the number of cloud-free observations for each pixel in each year.  The \textit{avg\_vis} product is the annual average light without any filtering. The \textit{stable\_lights.avg\_vis} product represents the annual average product in which ephemeral light and background noise have been eliminated. The last type was adopted in this study because its main light sources are built-up areas, which are the focus of most of the related studies. The dimension of a global DMSP-OLS image is (16801,43201), and its values range from 0 to 63. 

Due to lack of onboard calibration and the orbital drift, DMSP-OLS data is inconsistent between years and between satellites \citep{PANDEY201767,rs9060637,en20300595,jeswani2019evaluation}. It is necessary to perform an effective inter-calibration to enhance the continuity from 1992 to the present. We used the spatial and temporal variation coefficients to select calibration fields. The details are presented in Appendix \ref{AppendixA}.

The Earth Observation Group also released the NPP-VIIRS dataset\footnote{https://eogdata.mines.edu/products/vnl/} from April 2012, as shown in green. The overpass time of NPP-VIIRS is 01:30 after midnight \citep{rs5126717}. There are several versions of the dataset, and the \textit{average-masked} one was adopted in this work.  This version is the annual average radiance product in which biomass burning, aurora, and most of the background noise were removed. The dimension of a global NPP-VIIRS image is (33601, 86401), which is twice as large as that of a global DMSP-OLS image when the units digit is ignored. The unit of its value is $\emph{nanoWatts}/cm^2/Sr$. 

\section{DeepNTL model}
    In the previous models, multi-source data was reconstructed into NPP-VIIRS data using the direct mapping approach. However, this approach led to several issues, including (a) The need to discard parts of DMSP-OLS historical archives to align the times of different variables; (b) Insufficient accuracy; and (c) Degraded generalization ability in untrained years. Therefore, to fully exploit the historical archives of the DMSP-OLS, these auxiliary variables should be circumvented. In addition, to improve the accuracy and maintain a good generalization ability in untrained years, the direct mapping approach should also be circumvented.

We proposed a new super-resolution framework, learning annual difference, for NTL reconstruction. If the NPP-VIIRS image of a certain year can be selected as a reference, and the super-resolution reconstruction for the target year can be regarded as a brightness change on the reference image, then the super-resolution problem can be simplified. Moreover, the basis for the brightness change on the NPP-VIIRS image can be learned from the brightness difference in the DMSP-OLS images of different years. In this way, the above direct mapping approach and auxiliary variables can be avoided.

${x}^\prime$ is considered as the reference year, and a DMSP-OLS image of year ${x}^\prime$ is denoted as $\emph{DMSP}_{x^\prime y^\prime}$ whose satellite code is ${y}^\prime$. $x$ is considered as the target year, and a DMSP-OLS image of year $x$ is represented as $\emph{DMSP}_{xy}$ whose satellite code is $y$.  $\emph{DMSP}_{x^\prime y^\prime}$ and $\emph{DMSP}_{xy}$ are located in the same place. Function $F_1$ is the feature extractor of DMSP-OLS images. The feature difference between $\emph{DMSP}_{x^\prime y^\prime}$ and $\emph{DMSP}_{xy}$ represents the annual difference of DMSP-OLS, as expressed in Eq.\ref{eq1}:
\begin{equation}
    \label{eq1}
    F_2\left(\emph{DMSP}_{x^\prime y^\prime},\emph{DMSP}_{x y} \right)=F_1\left(\emph{DMSP}_{x^\prime y^\prime}\right)-F_1\left(\emph{DMSP}_{xy}\right),
\end{equation}
where $F_2\left(\emph{DMSP}_{x^\prime y^\prime},\emph{DMSP}_{x y} \right)$ is the annual difference of DMSP-OLS;  the first item in the right is the feature of $\emph{DMSP}_{x^\prime y^\prime}$; the second item in the right is the feature of $\emph{DMSP}_{xy}$.

Similarly, $\emph{VIIRS}_{x^\prime}$ denotes the NPP-VIIRS image of the reference year $x^\prime$.  It covers the same area as $\emph{DMSP}_{x^\prime y^\prime}$.  $\emph{VIIRS}_x$, also in the same place, is the NPP-VIIRS image of the target year $x$. Function $F_3$ is the feature extractor of the NPP-VIIRS images. The feature difference between $\emph{VIIRS}_{x^\prime}$ and $\emph{VIIRS}_x$ is the annual difference of NPP-VIIRS, as expressed in Eq.\ref{eq2}:
\begin{equation}
    \label{eq2}
    F_4\left(\emph{VIIRS}_{x^\prime},\emph{VIIRS}_x \right)=F_3\left(\emph{VIIRS}_{x^\prime}\right)-F_3\left(\emph{VIIRS}_x\right),
\end{equation}
where $F_4\left(\emph{VIIRS}_{x^\prime},\emph{VIIRS}_x \right)$ is the annual difference of NPP-VIIRS; the first item in the right is the feature of $\emph{VIIRS}_{x^\prime}$; the second item in the right is the feature of $\emph{VIIRS}_x$.

Eq.\ref{eq2} is reshaped as Eq.\ref{eq3}. If there is a transform function $H$, as shown in Eq.\ref{eq4}, which can transform $F_2\left(\emph{DMSP}_{x^\prime y^\prime},\emph{DMSP}_{x y} \right)$ into $F_4\left(\emph{VIIRS}_{x^\prime},\emph{VIIRS}_x \right)$, then we can change Eq.\ref{eq3} into Eq. \ref{eq5}. The new equation shows that with the help of $H$, we can combine the NPP-VIIRS feature $F_3\left(\emph{VIIRS}_{x^\prime}\right)$ of the reference year and the annual difference of DMSP-OLS $F_2\left(\emph{DMSP}_{x^\prime y^\prime},\emph{DMSP}_{x y} \right)$ to obtain the NPP-VIIRS feature $F_3\left(\emph{VIIRS}_x\right)$ of the target year. Then, Eq.\ref{eq1} is included in Eq.\ref{eq5} to get Eq.\ref{eq6}. 
\begin{equation}
    \label{eq3}
    F_3\left(\emph{VIIRS}_x\right)=F_3\left(\emph{VIIRS}_{x^\prime}\right)-F_4\left(\emph{VIIRS}_{x^\prime},\emph{VIIRS}_x \right)
\end{equation}
\begin{equation}
    \label{eq4}
    F_4\left(\emph{VIIRS}_{x^\prime},\emph{VIIRS}_x \right)=H\left(F_2\left(\emph{DMSP}_{x^\prime y^\prime},\emph{DMSP}_{x y} \right)\right)
\end{equation}
\begin{equation}
    \label{eq5}
    F_3\left(\emph{VIIRS}_x\right)=F_3\left(\emph{VIIRS}_{x^\prime}\right)-H\left(F_2\left(\emph{DMSP}_{x^\prime y^\prime},\emph{DMSP}_{x y} \right)\right)
\end{equation}
\begin{equation}
    \label{eq6}
    F_3\left(\emph{VIIRS}_x\right)=F_3\left(\emph{VIIRS}_{x^\prime}\right)-H\left(F_1\left(\emph{DMSP}_{x^\prime y^\prime}\right)-F_1\left(\emph{DMSP}_{xy}\right)\right)
\end{equation}

To obtain the NPP-VIIRS image instead of the feature of the target year, a reconstruction function $G$ is needed, as shown in Eq.\ref{eq7}. This is the initial prototype of our super-resolution model.
\begin{equation}
    \label{eq7}
    \emph{VIIRS}_x=G\left(F_3\left(\emph{VIIRS}_{x^\prime}\right)-H\left(F_1\left(\emph{DMSP}_{x^\prime y^\prime}\right)-F_1\left(\emph{DMSP}_{xy}\right)\right)\right)
\end{equation}

The minus sign in Eq.\ref{eq7} represents the ideal linear scenario, however, the actual scenario may be more complicated. To better reflect the complex and non-linear nature of the neural network models, we assume a new function $H^\ast$ that learns to capture the annual difference of DMSP-OLS, and then transforms it into that of NPP-VIIRS. Similarly, we assume a new function $G^\ast$ that learns to deduct the annual difference of NPP-VIIRS from its features in the reference year to get its features in the target year, and then reconstruct the features into an image. In this way, Eq.\ref{eq8} is obtained and shown below. This is the final prototype of our model. Of deep learning models, $G^\ast$, $F_3$, $H^\ast$, and $F_1$ are different modules, and their parameters can be learned through data.
\begin{equation}
    \label{eq8}
    \emph{VIIRS}_x=G^\ast\left(F_3\left(\emph{VIIRS}_{x^\prime}\right),H^\ast\left(F_1\left(\emph{DMSP}_{x^\prime y^\prime}\right),F_1\left(\emph{DMSP}_{xy}\right)\right)\right)
\end{equation}

\begin{figure}[!htbp]
\centering
    \includegraphics[width=1.0\textwidth]{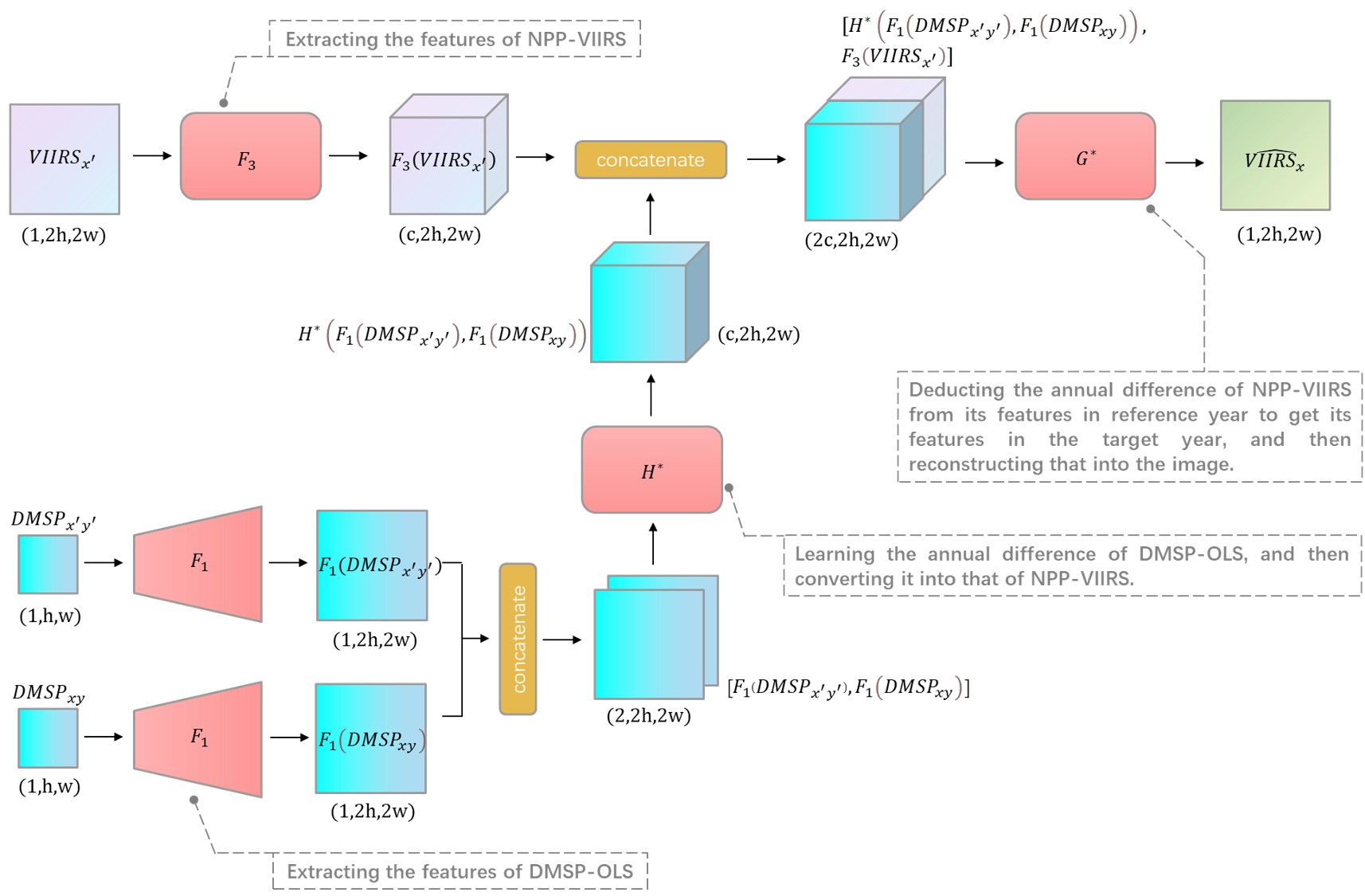}        
\caption{DeepNTL model. $\emph{DMSP}_{x^\prime y^\prime}$ denotes a DMSP-OLS image of the reference year $x^\prime$ whose satellite code is $y^\prime$. $\emph{DMSP}_{xy}$ is a DMSP-OLS image of the target year $x$ whose satellite code is $y$. $\emph{VIIRS}_{x^\prime}$ represents a NPP-VIIRS image of the reference year. $\emph{VIIRS}_{x}$ indicates a NPP-VIIRS image of the target year. $F_1$ is the module that extracts the features of DMSP-OLS images. $F_3$ is the module that extracts the features of the NPP-VIIRS image. $H^\ast$ is the module that learns the annual difference. $G^\ast$ is the module that reconstructs the NPP-VIIRS features into an image. The height and width of the DMSP-OLS image are denoted as h and w.}\label{fig02}
\end{figure}

According to Eq.\ref{eq8}, a new deep learning model, DeepNTL, was proposed, as shown in Fig.\ref{fig02}. The sizes of $\emph{DMSP}_{x^\prime y^\prime}$ and $\emph{DMSP}_{xy}$ are $(1, h, w)$, which indicates that the channel number, height, and width are 1, $h$, and $w$, respectively. When extracting the features of DMSP-OLS images, module $F_1$ doubles the heights and widths of images, because NPP-VIIRS images are twice as wide and high as that of DMSP-OLS images. Therefore, the sizes of the extracted features $F_1\left(\emph{DMSP}_{x^\prime y^\prime}\right)$ and $F_1\left(\emph{DMSP}_{x y}\right)$ are $(1, 2h, 2w)$.

$F_1\left(\emph{DMSP}_{x^\prime y^\prime}\right)$ and $F_1\left(\emph{DMSP}_{xy}\right)$ are concatenated along the channel dimension to obtain a tensor with a size of $(2,2h,2w)$. When learning the annual difference of DMSP-OLS and transforming it into that of NPP-VIIRS, the module $H^\ast$ increases the tensor’s channel number to $c$.

The size of $\emph{VIIRS}_{x^\prime}$ is $(1,2h,2w)$. The module $F_3$ increases its channel number to $c$ when extracting its features. Therefore, the tensor $F_3\left(\emph{VIIRS}_{x^\prime}\right)$ with a size of $(c,2h,2w)$ is created by this module.

Then, the NPP-VIIRS feature $F_3\left(\emph{VIIRS}_{x^\prime}\right)$ of the reference year and its annual difference \\$H^\ast\left(F_1\left(\emph{DMSP}_{x^\prime y^\prime}\right), F_1\left(\emph{DMSP}_{xy}\right)\right)$ are concatenated to obtain a new tensor whose size is $(2c,2h,2w)$. The module $G^\ast$ decreases the channel number from 2c to 1 during the reconstruction. The final output $\emph{VIIRS}_x$ with the size of $(1,2h,2w)$ is the NPP-VIIRS image of the target year.

In this study, the modified Residual Network (ResNet) was used as the module $H^\ast$, $F_3$ and $G^\ast$ \citep{He_2016_CVPR}. The Residual Channel Attention Network (RCAN) was used as the module $F_1$ \citep{Zhang_2018_ECCV}. The architectures and hyperparameters of ResNet and RCAN are described in \hyperref[AppendixB]{Appendix B} further.

\section{Implementation and Experiments}
    \subsection{Datasets}

    \begin{figure*}[!htbp]
        \centering
        \includegraphics[width=1.0\textwidth]{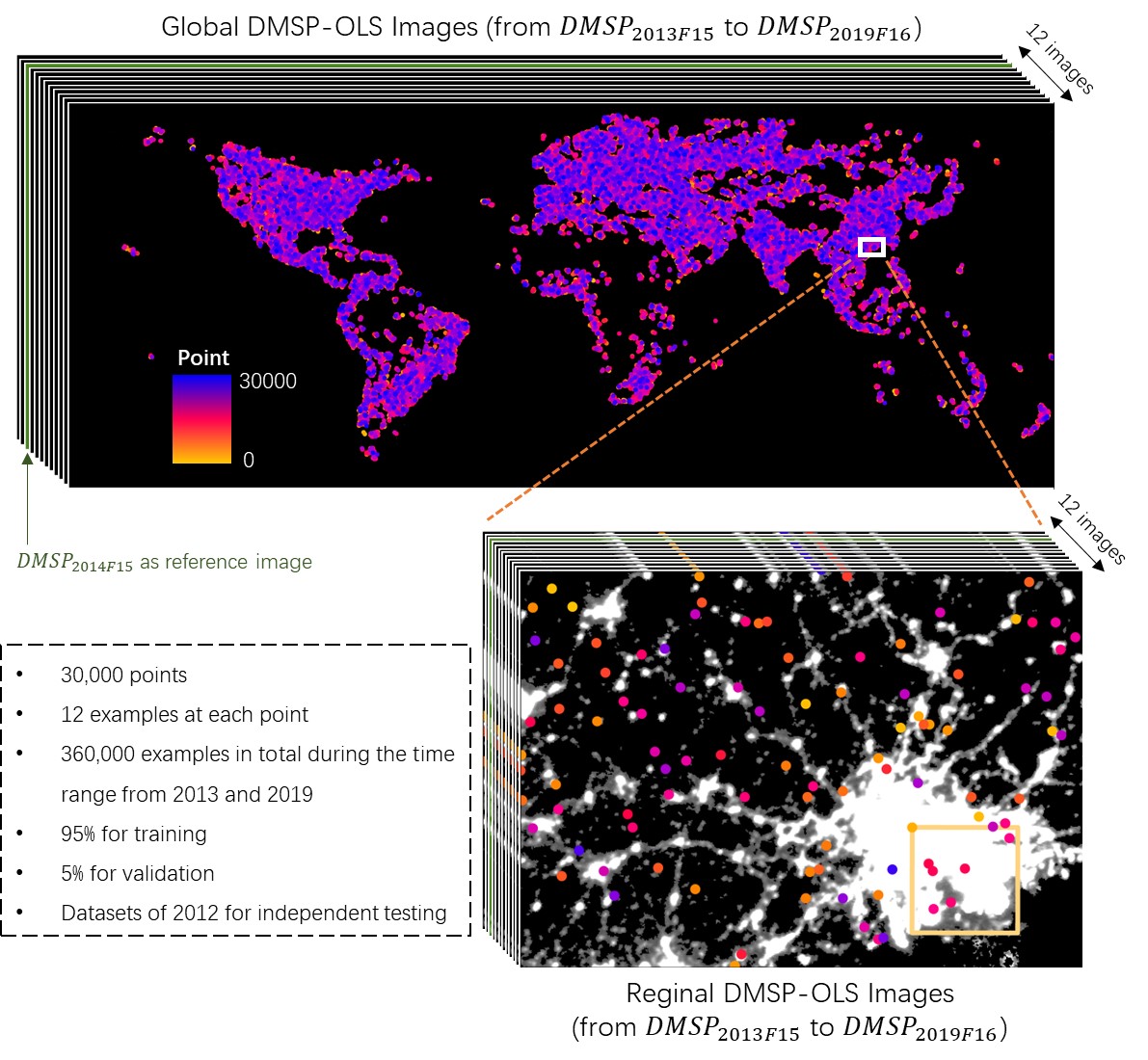}
        \caption{Global random sampling. The upper part shows the global DMSP-OLS images from $\emph{DMSP}_{2013F15}$ to $\emph{DMSP}_{2019F16}$ and randomly sampled points. The bottom right part shows the images and points in Guangdong-Hong Kong-Macao Greater Bay Area of China. The yellow box represents a tile.}\label{fig03}
    \end{figure*}

    The years from 2012 to 2019 are the intersecting times of DMSP-OLS and NPP-VIIRS. The DMSP-OLS products in the intersecting years include $\emph{DMSP}_{2012F18}$, $\emph{DMSP}_{2013F15}$, $\emph{DMSP}_{2013F18}$, $\emph{DMSP}_{2014F15}$, $\emph{DMSP}_{2015F15}$, $\emph{DMSP}_{2016F15}$, $\emph{DMSP}_{2016F16}$, $\emph{DMSP}_{2017F15}$, $\emph{DMSP}_{2017F16}$, $\emph{DMSP}_{2018F15}$, $\emph{DMSP}_{2018F16}$, $\emph{DMSP}_{2019F15}$, and $\emph{DMSP}_{2019F16}$. The products of NPP-VIIRS during this period include $\emph{VIIRS}_{2012}$, $\emph{VIIRS}_{2013}$, $\emph{VIIRS}_{2014}$, $\emph{VIIRS}_{2015}$, $\emph{VIIRS}_{2016}$, $\emph{VIIRS}_{2017}$, $\emph{VIIRS}_{2018}$, and  $\emph{VIIRS}_{2019}$. To eliminate the background noise in NPP-VIIRS data, the pixels less than 0.5 were assigned as 0.0.  In addition, a few pixels have abnormally high values in NPP-VIIRS images. These abnormal values may result from some unstable factors such as flames from the burning of natural gas, and hence, they were usually replaced using specific values in previous studies.  In this work, it was found that the value corresponding to 99.99\% quantile is 496 $\emph{nanoWatts}/{cm}^2/Sr$ among all lit pixels in all global NPP-VIIRS images. Therefore, the pixels greater than 496 $\emph{nanoWatts}/{cm}^2/Sr$ were replaced by it.

    The reference year $x^\prime$ was selected between 2013 and 2019. The selection was based on two considerations: (a) For the super-resolution reconstruction between 1992 and 2011, it is beneficial to select a reference year close to this period; (b) The model should not only learn the light change with the increase of year, but also learn that with the decrease of year.  Therefore, 2014 was determined as the reference year, $\emph{DMSP}_{2014F15}$ was used as $\emph{DMSP}_{x^\prime y^\prime}$, and $\emph{VIIRS}_{2014}$ was used as $\emph{VIIRS}_{x^\prime}$.

    The global images are too large to be directly input into the model. Therefore, it is necessary to extract tiles from the global images. The size of the DMSP-OLS tile is $(1,128,128)$, and that of the NPP-VIIRS tile is $(1,256,256)$. The strategy of global random sampling was used to determine the upper left points of tiles, as shown in Fig.\ref{fig03}. This strategy can increase the number of tiles, and thus enables the model to learn more possible situations. We generated 30,000 random points all over the world. For each point, the percentages of lit pixels in its $\emph{DMSP}_{2014F15}$ tile and the $\emph{VIIRS}_{2014}$ counterpart were both larger than $1\%$.

    The selected points were used to extract tiles from $\emph{DMSP}_{2013F15}$ to $\emph{DMSP}_{2019F16}$, and hence, each point extracted 12 DMSP-OLS tiles. Each DMSP-OLS tile was combined with the NPP-VIIRS counterpart in the same position and same year to form an example. Therefore, 360,000 examples were created in total. $95\%$ examples among them were used for training, and $5\%$ examples were used for validation.  The datasets of 2012 were used for independent testing.

\subsection{Loss Function}
    
    $L_1$ loss was used in this study. A set is denoted as $\left\{{\emph{VIIRS}_x^n,\widehat{\emph{VIIRS}_x^n}}\right\}_{n=1}^N$, which contains $N$ reconstructed images $\emph{VIIRS}_x^n$ and ground truth images $\widehat{\emph{VIIRS}_x^n}$. The loss function can be expressed as Eq.\ref{eq9}:
    \begin{equation}
        \label{eq9}
        L(\Theta)=\frac{1}{N} \sum_{n=1}^N\left\|\emph{VIIRS}_x^n-\widehat{\emph{VIIRS}_x^n}\right\|_1,
    \end{equation}
    where $L$ is the loss function; $\Theta$ is the set of parameters of the model; $N$ represents the number of examples; $n$ is the index of each example.

\subsection{Evaluation Metrics}

    Pearson correlation coefficient ($r$), peak signal-to-noise ratio ($\emph{PSNR}$), and structural similarity measure ($\emph{SSIM}$) were used to evaluate the consistency between a super-resolution image (SR) and  a ground truth NPP-VIIRS image (GT). $r$ can be expressed as follows:        
    \begin{equation}
        \label{eq10}
        r=\frac{\sum_{i=1}^I\left(p_i-\mu_{GT}\right)\left(\widehat{p}_i-\mu_{SR}\right)}{\sqrt{\sum_{i=1}^I\left(p_i-\mu_{GT}\right)^2} \sqrt{\sum_{i=1}^I\left(\widehat{p}_i-\mu_{SR}\right)^2}},
    \end{equation}
    \begin{equation}
        \label{eq11}
        \mu_{GT}=\frac{1}{I} \sum_{i=1}^I p_i,
    \end{equation}
    \begin{equation}
        \label{eq12}
        \mu_{SR}=\frac{1}{I} \sum_{i=1}^I \widehat{p}_l,
    \end{equation}
    where $\mu_{GT}$ is the average pixel value of a GT image; $\mu_{SR}$ represents the average pixel value of the corresponding SR image; $I$ is the number of pixels, and $i$ is the pixel index; $p_i$ and $\widehat{p_i}$ represent the $i$ th pixel value in GT and SR images, respectively. $r$ ranges from −1 to 1, and a larger value indicates a stronger correlation. PSNR is expressed as follows:
    \begin{equation}
        \label{eq13}
        \emph{PSNR}=10 \log_{10}\left(\frac{\emph{MAX}^2}{\emph{MSE}}\right),
    \end{equation}
    \begin{equation}
        \label{eq14}
        \emph{MSE}=\frac{1}{I} \sum_{i=1}^I\left(p_i-\widehat{p}_l\right)^2,
    \end{equation}
    where $\emph{MSE}$ represents the mean square error; $\emph{MAX}$ is the possible maximum value. A higher value of $\emph{PSNR}$ represents a smaller difference between the GT and SR images. Another metric, $\emph{SSIM}$, is shown as follows:
    \begin{equation}
        \label{eq15}
        \emph{SSIM}=\frac{\left(2 \mu_{GT} \mu_{SR}+(0.01\emph{MAX})^2\right)\left(2 \sigma_{GT \_SR}+(0.03\emph{MAX})^2\right)}{\left(\mu_{GT^2}+\mu_{SR^2}+(0.01 \emph{MAX})^2\right)\left(\sigma_{GT^2}+\sigma_{SR^2}+(0.03\emph{MAX})^2\right)},
    \end{equation}
    \begin{equation}
        \label{eq16}
        \sigma_{GT}=\sqrt[2]{\frac{1}{I} \sum_{i=1}^I\left(p_i-\mu_{G T}\right)^2},
    \end{equation}
    \begin{equation}
        \label{eq17}
        \sigma_{SR}=\sqrt[2]{\frac{1}{I} \sum_{i=1}^I\left(\widehat{p}_{\imath}-\mu_{SR}\right)^2},
    \end{equation}
    \begin{equation}
        \label{eq18}
        \sigma_{GT_{-}SR}=\frac{1}{I} \sum_{i=1}^{I}\left(p_{i}-\mu_{GT}\right)\left(\widehat{p}_{\imath}-\mu_{SR}\right),
    \end{equation}
    where $\sigma_{GT}$ and $\sigma_{SR}$ are the standard deviations of pixel values in the GT and SR images, respectively; $\sigma_{GT\_SR}$ is the covariance between the GT and SR images. The range of $\emph{SSIM}$ is between 0 and 1, and a large value of $\emph{SSIM}$ indicates a higher consistency between GT and SR images.

\subsection{Training and Inference}

    The model's parameters were randomly initialized. The Adam algorithm was used as the optimizer. The initial learning rate was set as 0.0001. The learning rate was reduced to its 95\% as long as the validation loss did not decrease for 3 epochs. We used 8 NVIDIA A10 graphic cards to train our DeepNTL parallelly for approximately three weeks. The batch size for each card was set as 4. During the training, each epoch took more than 4 h, and each card was fully utilized. During the inference, the global DMSP-OLS images from 1992–2019 were split into tiles, and then fed into the trained model. After all the super-resolution results were generated, they were re-organized as global images. 

\subsection{Baseline Models}

    The bilinear model is a commonly used, simple baseline model. The RCAN is a baseline model based on convolution operation. The SwinIR is another baseline model for image restoration based on Swin Transformer \citep{Liang_2021_ICCV}. Additionally, the AutoEncoder is the first model used to convert the DMSP-OLS images into the NPP-VIIRS images, and its productions are open-access, so the production of 2012 was used for comparison in this study. All of them belong to the direct mapping paradigm.

\section{Result and Discussion}
    \subsection{Evaluating visual consistency of super resolution}
    \begin{figure*}[!htbp]
        \centering
        \includegraphics[width=0.82\textwidth]{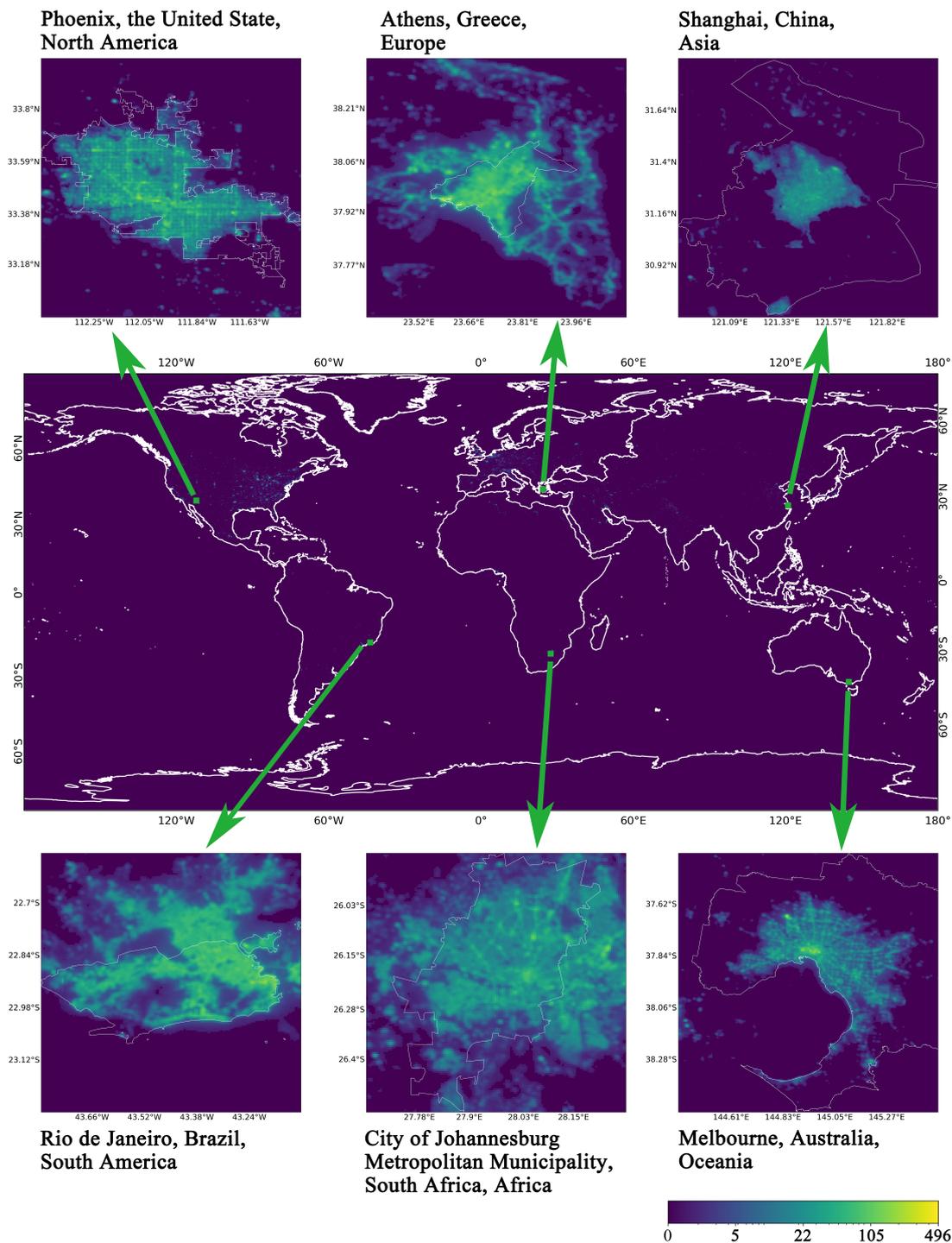}
        \caption{Global reconstructed NTL in 1992 using our DeepNTL model and its regional images of some cities.}\label{fig04}
    \end{figure*}
    Visual consistency was evaluated using image textures. Six big cities from different continents were used for the detailed analysis, including Shanghai in Asia, Melbourne in Oceania, Athens in Europe, Johannesburg in Africa, Phoenix in North America, and Rio de Janeiro in South America. Figure \ref{fig04} shows the global reconstructed NTL image in 1992 using our DeepNTL model as well as its regional images of these cities. Among these cities, Shanghai, Johannesburg, and Rio de Janeiro are located in developing countries, and Phoenix, Athens, and Melbourne are located in developed countries. Johannesburg and Phoenix are interior cities and the others are coastal cities. These cities have different socioeconomic backgrounds and geographical conditions, and hence are representative for the evaluation. Additionally, the NTL images of these cities are bright with distinctive textures, which facilitates comparison of the models' performances.

    \begin{figure*}[!htbp]
        \centering
        \includegraphics[width=0.80\textwidth]{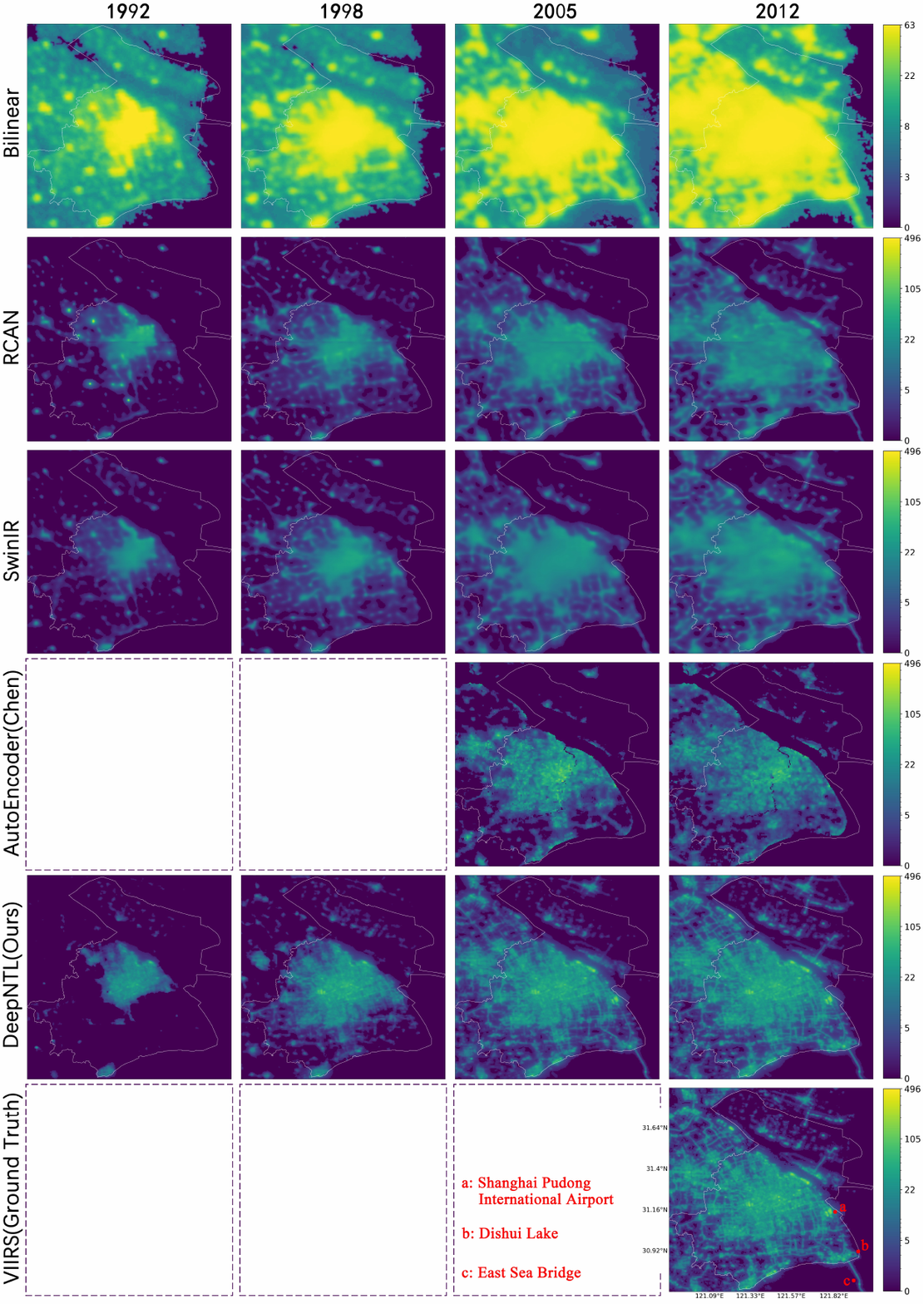}
        \caption{Reconstructed and GT images of Shanghai. (For better perception, readers are recommended to zoom in on the web version of this figure.)}\label{fig05}
    \end{figure*}

    \begin{figure*}[!htbp]
        \centering
        \includegraphics[width=0.80\textwidth]{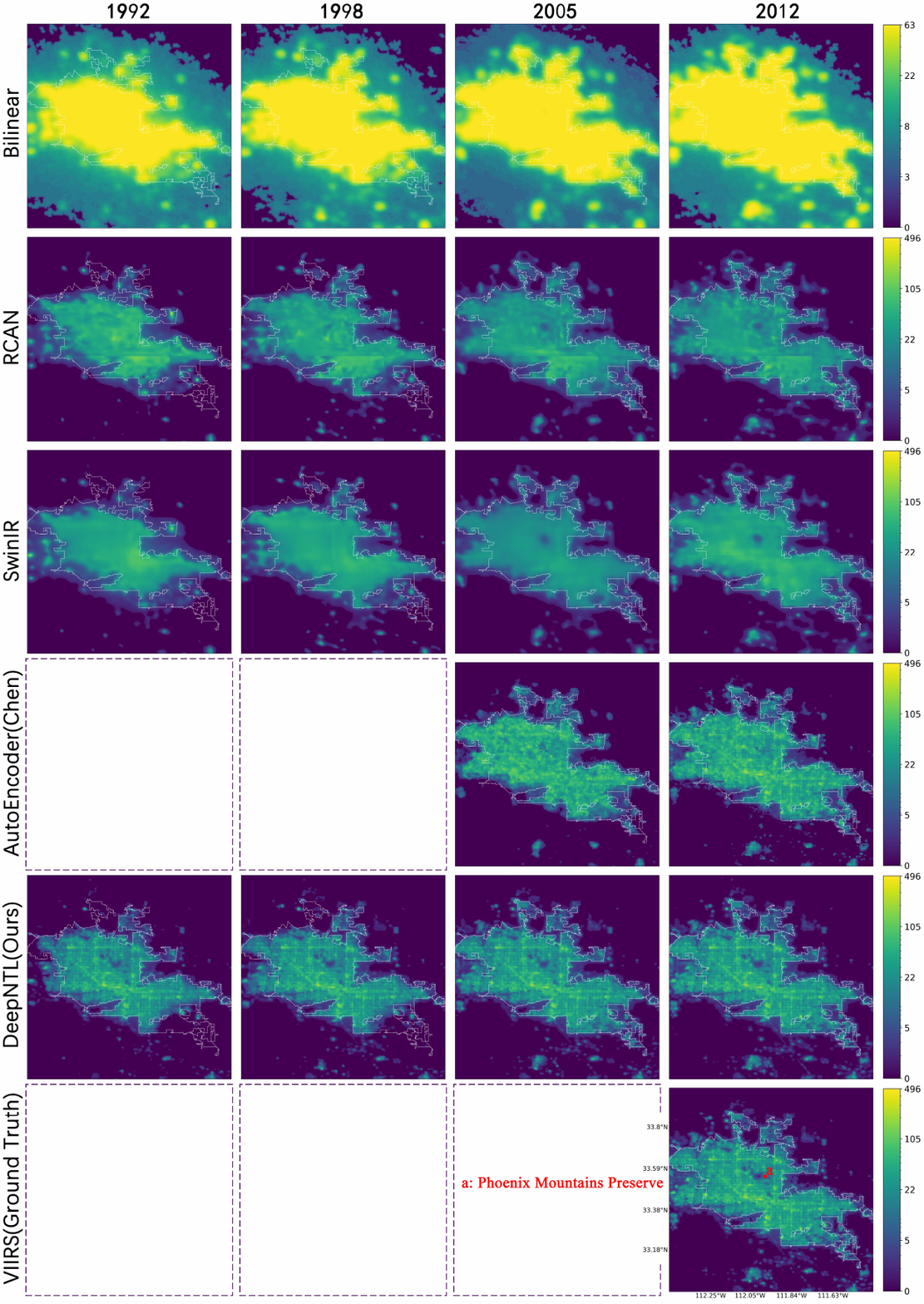}
        \caption{Reconstructed and GT images of Phoenix. (For better perception, readers are recommended to zoom in on the web version of this figure.)}\label{fig06}
    \end{figure*}
    Shanghai borders the East Sea to the east, Hangzhou Bay to the south, inland to the west, and the Yangtze River estuary to the north. Figure \ref{fig05} shows the reconstructed and GT images in Shanghai from 1992–2012. Columns and rows represent different years and models, respectively. The GT image is presented in the last row. The images produced by the bilinear model, displayed in the first row, suffer from a couple of issues. Firstly, its images are blurry, making it very difficult to distinguish details such as streets or buildings. Secondly, the model's results inherit the overglow effect from the original DMSP-OLS images. As an example, the vast sea area to the east of Shanghai is erroneously illuminated Since 2005, even though there is no stable human activity on the sea. Additionally, the images suffer from a saturation problem, resulting in no brightness variation in the urban center. The second row presents the images of the RCAN model, which displays some roads outside the urban center. The model partially alleviates the overglow effect by maintaining the darkness of the eastern sea. It also overcomes the saturation problem to exhibit some brightness variations in the urban center. However, its images are still indistinct within the urban center. The images produced by SwinIR, shown in the third row, share similar characteristics with RCAN, although they are slightly smoother. The AutoEncoder's images are displayed in the fourth row. The model successfully overcomes the overglow effect and the saturation problem, because it does not illuminate the pixels on the East Sea and shows some brightness variations in the urban center. Compared to the previous models, AutoEncoder enhances textures a lot. However, its textures are not entirely consistent with the GT image, making it difficult to discern streets and buildings. Additionally, as the model's input is the combination of DMSP-OLS image and MODIS EVI, and the latter starts from 2000, it cannot reconstruct the images in 1992 and 1998. Furthermore, the images inherit the water mask from MODIS EVI, resulting in some bright pixels in the urban center being set to 0. The water masks appear as south-north dark curves between 121.33°E and 121.57°E in the images produced by AutoEncoder. Finally, the fifth row shows the reconstructed images of the DeepNTL model. Its images successfully overcome the overglow effect and saturation problem, while also being free of any water mask, making them complete. As it does not require any auxiliary data, the entire DMSP-OLS archives since 1992 can be fully utilized. Moreover, its images are clear enough to display streets and other details in all years, and the image in 2012 is highly consistent with the GT image.

    The unprecedented fine-grained and long-term NTL analysis can now be supported by DeepNTL. For example, point a in the GT image of Figure \ref{fig05} indicates the Shanghai Pudong International Airport, the largest hub airport in East China, that was completed in 1999. Its lights are very bright and conspicuous. According to the DeepNTL images, this spot did not show up in 1992 and 1998, and became increasingly clear in 2005 and 2012. Hence, what the DeepNTL shows matches the facts. Furthermore, point b indicates the Dishui Lake, which is an artificial, circular lake.It was mostly completed in 2003, with ongoing refinements since then. It appears as a small ring on the GT image, which is formed by street lamps surrounding the lake. On the DeepNTL images, this small ring did not exist before 1998; by 2005 it had appeared and became evident in 2012. This is also consistent with the facts. Lastly, point c indicates the East Sea Bridge, a sea-crossing bridge connecting Shanghai with Zhoushan in the south, that was put into service in 2005. On the DeepNTL images, the bridge did not show prior to 1998. It appeared after 2005. This is again in line with the facts. Such long-time analysis for individual facilities is only possible with DeepNTL. In contrast, other models cannot show these details at all. DeepNTL has the same advantages in the analysis of Athens and Rio de Janeiro. To save article space, the images of these two cities are presented in Figure C1 and C2 of \hyperref[AppendixC]{Appendix C}. 

    Another advantage of DeepNTL is that it is able to maintain a strong generalization ability for earlier years that are far from the training years.  As depicted in Figure \ref{fig06}, the GT image of Phoenix shows a dark polygon area near point a that represents the Phoenix Mountain Preserve. It is much darker than the surrounding area as it has maintained its original ecology over the years.  The images of RCAN and SwinIR were able to show the dark mountain reserve in 2005 and 2012, but it disappeared from their images of 1992 and 1998. This is due to the decline in the generalization ability of the two models when they are applied in the early years. In addition, the AutoEncoder image of 2012 was able to display the dot-matrix-like block layout, which is similar to that of the GT image. However, in 2005, its images became blurred and were unable to show the block layout clearly. This, too, is caused by the decline in the generalization ability of the model. Notably, DeepNTL can clearly display the mountain reserve every year and maintain the dot-matrix-like block layout over the years. Incidentally, this city expands in the southeast direction. This advantage can also be found in Melbourne and Johannesburg. To save space in the article, the images of these two cities are presented in Figure C3 and C4 of \hyperref[AppendixC]{Appendix C}.

    Overall, DeepNTL has the strongest visual consistency when compared to other models, its advantages include: (a) It can reconstruct DMSP-OLS images of all years; (b) The reconstructed images are clear, and the textures are very close to that of GT images; (c) It maintains good generalization ability for earlier years and can accurately detect annual changes for individual facilities.
    
\subsection{Evaluating statistical consistency of super resolution} \label{Evaluating_statistical_consistency_of_super_resolution}

    \subsubsection{Statistical consistency at urban scale} \label{Stat_consistency_at_urban_scale}

        \begin{figure*}[!htbp]
            \centering
            \includegraphics[width=0.82\textwidth]{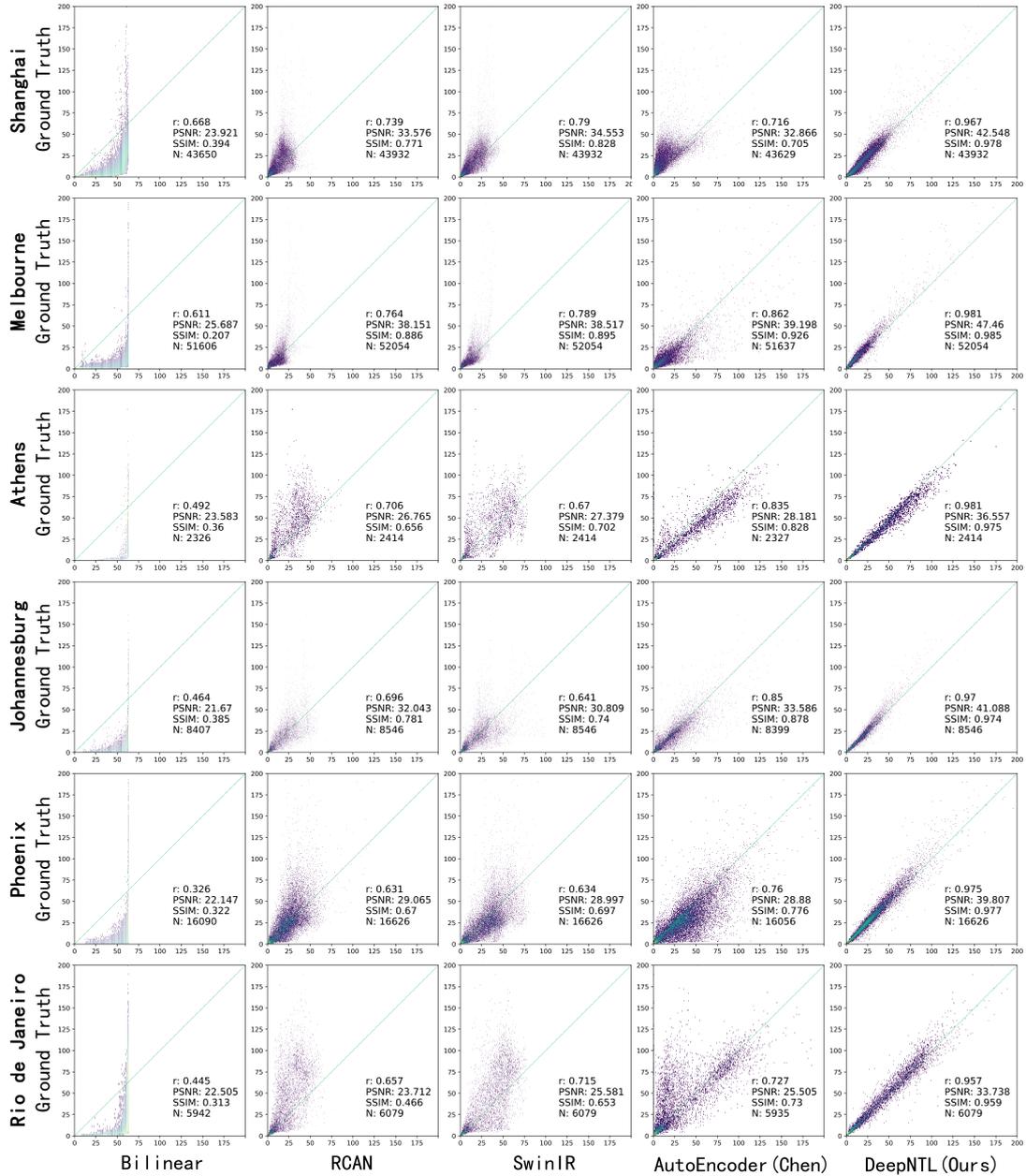}
            \caption{Pixel relationship at urban scale in 2012. Abscissa represents models' outputs, ordinate stands for GT. $r$ indicates correlation coefficient, $\emph{PSNR}$ represents peak signal-to-noise ratio, $\emph{SSIM}$ stands for structural similarity index measure, $N$ is pixel number. We zoom in to the subrange from 0 to 200 to show the most concentrated parts.}\label{fig07}
        \end{figure*}

        \begin{figure*}[!htbp]
            \centering
            \includegraphics[width=0.82\textwidth]{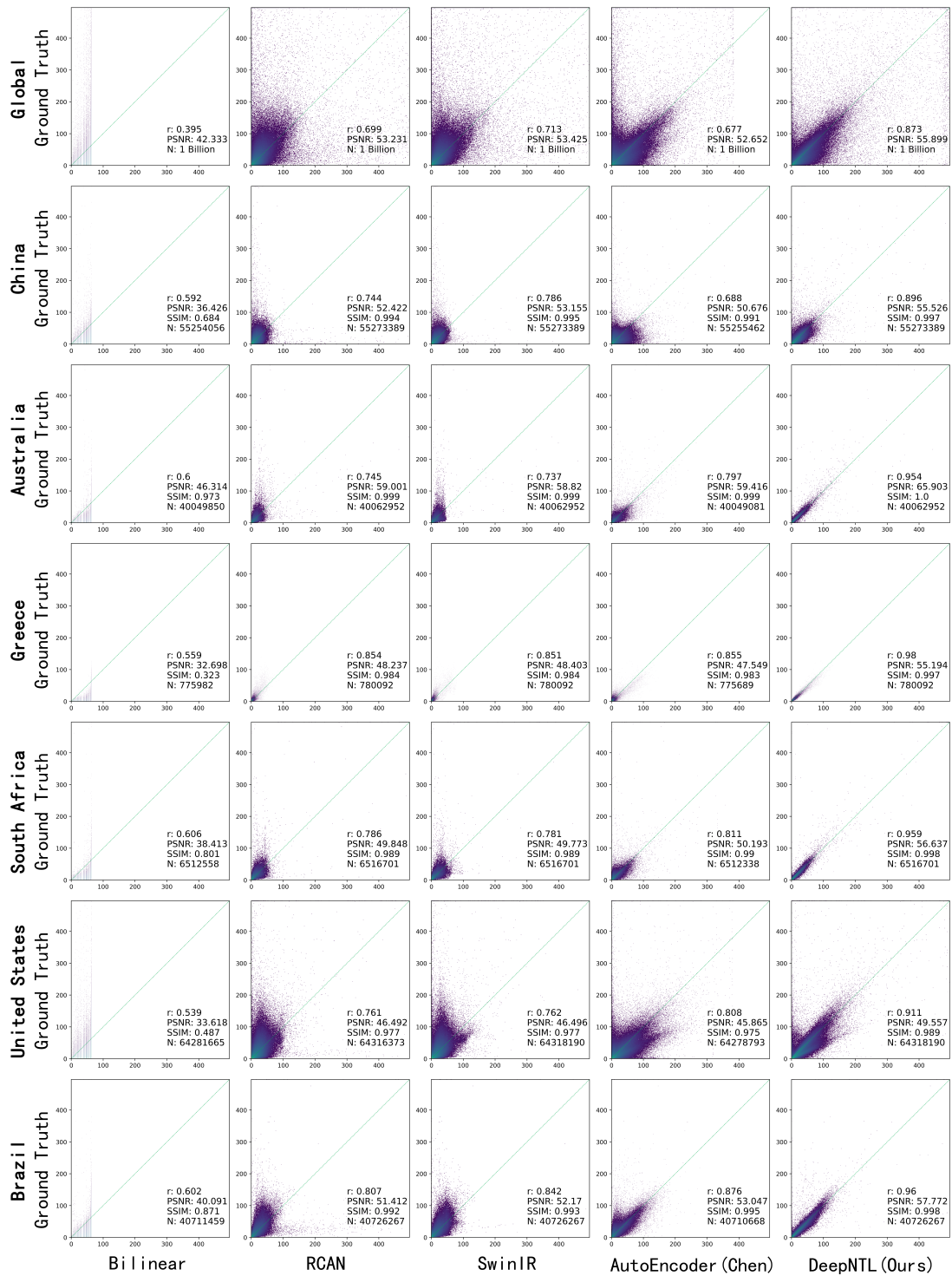}
            \caption{Pixel relationship at global and country scale in 2012. Abscissa represents models' outputs, ordinate stands for GT. $r$ indicates correlation coefficient, $\emph{PSNR}$ represents peak signal-to-noise ratio, $\emph{SSIM}$ stands for structural similarity index measure, $N$ is pixel number.}\label{fig08}
        \end{figure*}
      
        Figure \ref{fig07} shows the pixel relationship between the model outputs and GT images at an urban scale in 2012. All the pixels in these cities were used to calculate the evaluation metrics. To magnify the most concentrated parts, we zoomed in to the subrange between 0 and 200. The maximum value of the bilinear model is 63, as it does not make a substantive change to the DMSP-OLS images. The points of the bilinear model are concentrated on the abscissas due to the overglow effect. The points of RCAN, SwinIR, and AutoEncoder are more dispersed and have different degrees of overestimation or underestimation. In contrast, all the points of DeepNTL are concentrated along the diagonals.
        
        For each city, DeepNTL has the highest evaluation metrics.  For instance, in Shanghai, SwinIR is better than other direct mapping models, with $r$ of 0.79, $\emph{PSNR}$ of 34.553, and $\emph{SSIM}$ of 0.828.  Our DeepNTL outperforms SwinIR, whose $r$ is 0.967, $\emph{PSNR}$ is 42.548 and $\emph{SSIM}$ is 0.978. In Melbourne, $r$, $\emph{PSNR}$, and $\emph{SSIM}$ of AutoEncoder are 0.862, 39.198, and 0.926 respectively, and higher than those of other direct mapping models. DeepNTL is better than AutoEncoder, with $r$ of 0.981, $\emph{PSNR}$ of 47.46 and $\emph{SSIM}$ of 0.985. Similar results can be observed for Athens, Johannesburg, Phoenix, and Rio de Janeiro. The metrics of direct mapping models have unstable rankings in different cities. In contrast, the metrics of DeepNTL consistently rank first in each city and are significantly higher than those of other models. This is due to the ability of DeepNTL to learn the annual difference, which provides a distinct advantage.

    \subsubsection{Statistical consistency at global and country scale}

        The performances at global and country scales were evaluated by statistical consistency. Figure \ref{fig08} shows the pixel relationships between the output values of the models and GT values at these two scales in 2012. The first row represents the global comparison. One billion pixels were randomly selected from nearly three billion pixels worldwide for evaluation. Such a tremendous number of selected pixels is enough to illustrate the global performance. The metrics of the bilinear model are the lowest among all models. Its global $r$ is 0.395 and $\emph{PSNR}$ is 42.333, which indicates that there is a huge gap between the global images of bilinear model and GT. AutoEncoder improves the relationship significantly, with a $r$ value of 0.677 and $\emph{PSNR}$ of 52.652. Since the maximum value of AutoEncoder is less than 400, there is a blank area on the right of its global scatter diagram. RCAN is slightly better than AutoEncoder with a $r$ value of 0.699 and $\emph{PSNR}$ of 53.231, and its data range is consistent with that of GT. SwinIR is better than RCAN, with a $r$ value of 0.713 and $\emph{PSNR}$ of 53.425. Compared with the bilinear model, these direct mapping models see evident improvement at the global scale, but there is also obvious dispersion in their scatter diagrams. DeepNTL model has the best performance at the global scale with a $r$ value of 0.873 and $\emph{PSNR}$ of 55.899. Its data range is consistent with that of GT. The model's points are concentrated on the diagonal, and its dispersion degree is the lowest compared with that of other models.
    
        To evaluate statistical consistency at the country scale, China, Australia, Greece, South Africa, the United States, and Brazil were selected for comparison. All the pixels within each country were used to produce scatter diagrams. DeepNTL maintains the first place in each country. For example, in China, SwinIR performs better than other direct mapping models, with a $r$ value of 0.786, PSNR of 53.155, and SSIM of 0.995. However, DeepNTL outperforms SwinIR with a $r$ value of 0.896, PSNR of 55.526, and SSIM of 0.997. In Australia, AutoEncoder is better than other direct mapping models, with a $r$ value of 0.797, PSNR of 59.416, and SSIM of 0.999. But DeepNTL surpasses AutoEncoder with a $r$ value of 0.954, PSNR of 65.903, and SSIM of 1.0. In Greece, South Africa, the United States, and Brazil, DeepNTL still has the highest evaluation metrics.

\conclusions
    
NTL has played an irreplaceable role in analyzing human activity. DMSP-OLS provides the longest NTL historical archives from 1992. NPP-VIIRS is the new generation NTL sensor introduced in 2012, with a higher spatial resolution that gives it more application potential. The inconsistency between these two kinds of sensors results in the lack of a long-term, fine-grained NTL dataset. This problem has persisted without an effective solution for a long duration. We introduced a novel super-resolution framework based on the concept of learning annual difference. Using this framework, we developed a new model called DeepNTL, which is specifically designed for NTL data. We created a large dataset comprising 360,000 image examples to train our model. Through visual and statistical evaluations, we demonstrated that DeepNTL surpasses baseline models across multiple scales. In particular, DeepNTL is the only model that can accurately capture the dynamics of infrastructure such as airports and roads. 

Although our DeepNTL model and product prove significant advantages over other models on various fronts, the temporal resolution of our product is annual. Fortunately, the monthly DMSP-OLS and daily NPP-VIIRS have been released in recent years. Moreover, some new satellite datasets, e.g. the daily Luojia dataset since 2018 with 130 m spatial resolution, are also available. These new datasets can be used to produce long-term NTL datasets with higher spatiotemporal resolution in the future by our DeepNTL. Our future work also includes using DeepNTL products to conduct large-scale surveys of the long-term changes in global infrastructure, such as airports, roads, bridges, and so on.

For the first time, the long-term and fine-grained NTL observation becomes a reality. The DeepNTL product is a valuable extension of NPP-VIIRS, which provides reliable NTL data for earlier years. Further, it is open access to the public. Users can easily combine future NPP-VIIRS annual data with the DeepNTL product after removing background noise.

\dataavailability{The long-term and fine-grained nighttime light dataset is available at \url{https://doi.org/10.5281/zenodo.7859205}.}

\authorcontribution{JG developed the methodology and software, performed visualization, and drafted the manuscript; FZ conceived the study, performed validation and formal analysis, revised the manuscript, supervised the project, and administered the project; HZ performed investigation and provided resources, and revised the manuscript; BP performed validation and visualization, and revised the manuscript; LM performed formal analysis, and revised the manuscript.} 

\competinginterests{The authors declare that they have no competing interest.} 

\begin{acknowledgements}
The authors would like to thank Shanghai Qizhi Institute. They also would like to thank the Earth Observation Group of the Colorado School of Mines for providing nighttime light datasets.
In addition, they appreciate all the related studies. 
\end{acknowledgements}

\appendixfigures 
\appendixtables
\appendix
\newpage
\section{Inter-calibration for DMSP-OLS} \label{AppendixA} 
    \subsection{The selection of calibration fields}

    The homogeneity at both spatial and temporal dimensions matters for ideal calibration fields \citep{rs6032494,ZIBORDI2017122}. Hence, in addition to the spatial variation coefficient, we used a temporal variation coefficient to measure the temporal stability of the whole period.

    $\emph{GDMSP}_{xy}$ is a global DMSP-OLS image, in which $x$ is the year and $y$ is the satellite code. With each pixel as a center and three surrounding pixels as the kernel size, the spatial variation coefficient is calculated according to Eq.\ref{eqa1}:
    \begin{equation}
        \label{eqa1}
        vc_{i}^{s} = \frac{\sigma_{i}^{s}}{\mu_{i}^{s}},
    \end{equation}
    where $vc_{i}^{s}$ is the spatial variation coefficient of the kernel centered on pixel $i$; $\sigma_{i}^{s}$ represents the standard deviation of the kernel; $\mu_{i}^{s}$ is the average light within the kernel. In this way, the spatial variation coefficient image $\emph{SVC}_{xy}$ for each $\emph{GDMSP}_{xy}$ was obtained. Then, we sorted all the pixel values of all $\emph{SVC}_{xy}$, and took the commonly used $1/4$ quantile as the spatial threshold.  Pixels lower than this threshold are more uniform in space and were assigned 1.  In contrast, those higher than the threshold were assigned 0. After that, the spatial mask $\emph{SM}_{xy}$ for each $\emph{SVC}_{xy}$ was acquired.  Finally, the total spatial mask $\emph{TSM}$ was calculated by multiplying all the $\emph{SM}_{xy}$. The $\emph{TSM}$ represents the pixels with high spatial uniformity.

    All the $\emph{GDMSP}_{xy}$ were stacked together along the channel dimension to form a thick image $\emph{GDMSP}_{\emph{thick}}$.  The temporal variation coefficient was calculated using all channel values for each pixel, as shown in Eq.\ref{eqa2}:
    \begin{equation}
        \label{eqa2}
        vc_{i}^{t}= \frac{\sigma_{i}^{t}}{\mu_{i}^{t}},
    \end{equation}
    where $vc_{i}^{t}$ is the temporal variation coefficient for pixel $i$; $\sigma_{i}^{t}$ is the standard deviation of all channel values at this pixel; $\mu_{i}^{t}$ denotes the average, and thus, the temporal variation coefficient image $\emph{TVC}$ was obtained.  We also used the $1/4$ quantile as the temporal threshold, and then, binarized $\emph{TVC}$ in a similar way to $\emph{SVC}_{xy}$. After that, the temporal mask $\emph{TM}$ was created, which represents pixels having a high temporal stability. 
    
    In addition, each $\emph{GDMSP}_{xy}$ has some saturated pixels with a saturated value of 63.  The saturated pixels cannot represent true light values, hence, such pixels must be eliminated.  The unsaturated mask $\emph{USM}_{xy}$ for each $\emph{GDMSP}_{xy}$ was obtained by setting pixels equivalent to 63 as 0 and setting other pixels as 1. Subsequently, we multiplied all the $\emph{USM}_{xy}$ to obtain the total unsaturated mask $\emph{TUSM}$. 
    
    Finally, the calibration fields ($\emph{CF}$) were obtained by making an intersection between the total spatial mask ($\emph{TSM}$), temporal mask ($\emph{TM}$), and total unsaturated mask ($\emph{TUSM}$) as expressed in Eq.\ref{eqa3}:
    \begin{equation}
        \label{eqa3}
        \emph{CF}=\emph{TSM}\times \emph{TM}\times \emph{TUSM}
    \end{equation}

    \begin{figure*}[!htbp]
        \centering
        \includegraphics[width=1.0\textwidth]{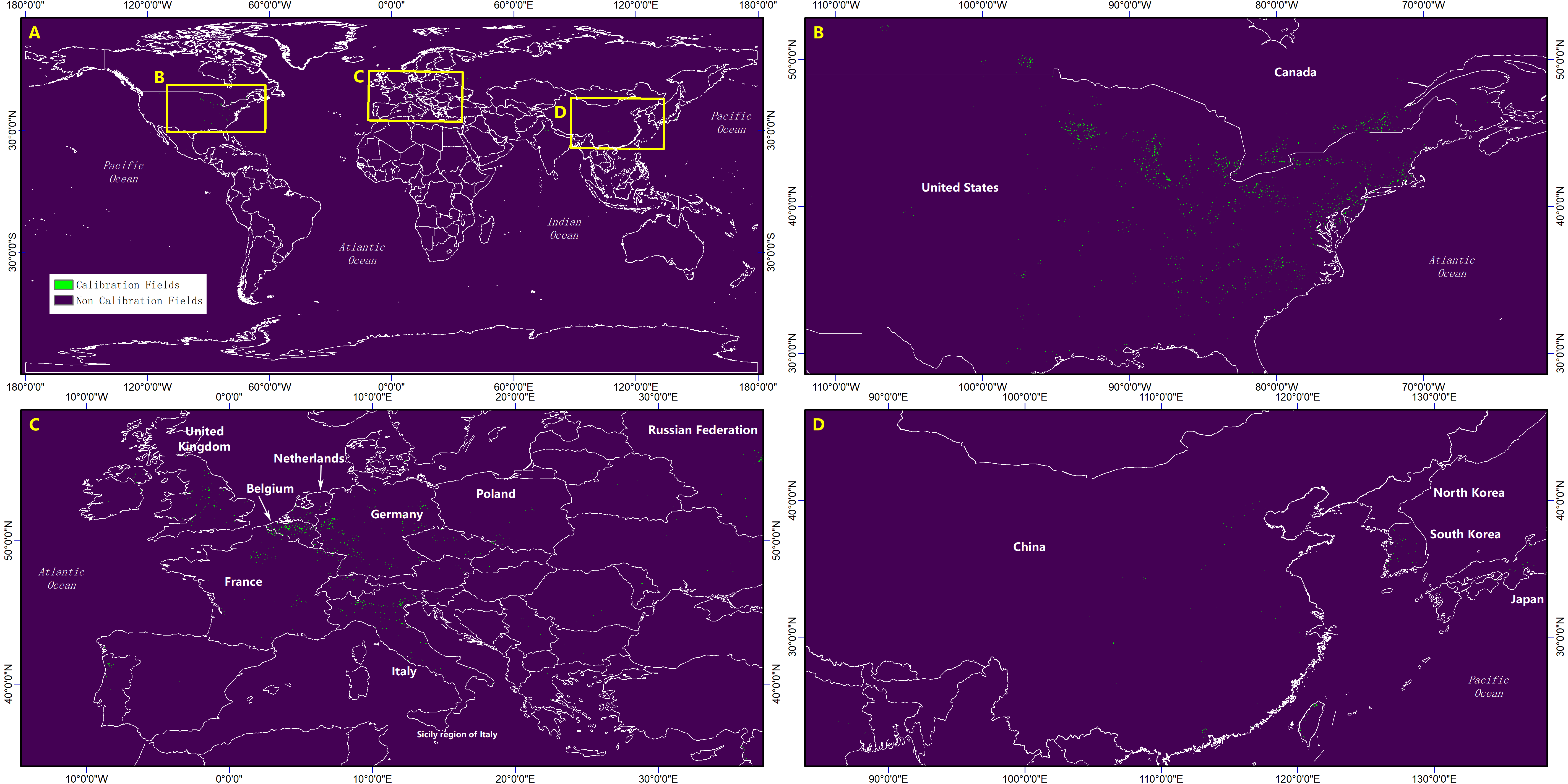}
        \caption{Distribution of calibration fields. (For better perception, readers are recommended to zoom in on the web version of this figure.)}\label{figA1}
    \end{figure*}
    
    Of a total of 725,820,001 pixels in a global DMSP-OLS image, 94,737 pixels constitute calibration fields, as shown in Figure \ref{figA1} (A). The calibration fields are mainly located in the Northern Hemisphere because most of the NTL is released by countries in this region. The calibration fields located in North America are shown in Figure \ref{figA1} (B). They are mainly distributed in eastern United States and southern Canada. These developed areas completed most of the infrastructure construction in the twentieth century, and hence, the NTL in these areas has remained temporally stable and spatially uniform during the past 30 years. Therefore, the calibration fields are densely distributed in these areas. As shown in Figure \ref{figA1} (C), the dense distribution of calibration fields is also found in western Europe. Previous studies chose the Sicily Island, Italy, as the calibration field. However, based on our method, it is found that only few pixels remain temporally stable and spatially uniform in Sicily. In fact, Belgium has the densest distribution of calibration fields in western Europe. In addition, the calibration fields are also distributed in the United Kingdom, France, Germany, and some other countries. As shown in Figure \ref{figA1} (D), China has a sparse distribution of calibration fields because it has a vast territory and its cities developed rapidly in recent decades.

    After the calibration fields were determined, based on Elvidge \citep{en20300595}, the quadratic polynomial function was used to correct the continuity of the DMSP-OLS. Fitted parameters and determination coefficients are presented in table \ref{fittedParameters}.

    \begin{table*}[htbp]
        \caption{Fitted parameters of the quadratic polynomial for DMSP-OLS continuity correction. Referencing Elvidge, $\emph{DMSP}_{1999F12}$ was used as the base image in the improvement of DMSP-OLS continuity.} 
        \begin{tabularx}{\textwidth}{XXXXXXXXXX}
        
        \toprule
        \textbf{Year} & \textbf{1992} & \textbf{1993} & \textbf{1994} & \textbf{1994} & \textbf{1995} & \textbf{1996} & \textbf{1997} & \textbf{1997} & \textbf{1998}\\
        \textbf{Satellite} & \textbf{F10} & \textbf{F10} & \textbf{F10} & \textbf{F12} & \textbf{F12} & \textbf{F12} & \textbf{F12} & \textbf{F14} & \textbf{F12} \\
        \midrule
        $a$ & -0.0107 & -0.0118 & -0.0075 & -0.0102 & -0.0062 & -0.0072 & -0.0041 & -0.0157 & -0.0033 \\
        $b$ & 1.6983 & 1.7771 & 1.4614 & 1.6623 & 1.4031 & 1.4873 & 1.2572 & 1.9777 & 1.1930 \\
        $c$ & -2.3134 & -2.8972 & -0.1966 & -2.5930 & -1.5095 & -2.0035 & -0.3701 & -2.1581 & -0.2953 \\
        $R^2$ & 0.9236 & 0.9311 & 0.9155 & 0.9627 & 0.9677 & 0.9712 & 0.9660 & 0.9661 & 0.9678 \\
        
        \toprule
        \textbf{Year} & \textbf{1998} & \textbf{1999} & \textbf{1999} & \textbf{2000} & \textbf{2000} & \textbf{2001} & \textbf{2001} & \textbf{2002} & \textbf{2002} \\
        \textbf{Satellite} & \textbf{F14} & \textbf{F12} & \textbf{F14} & \textbf{F14} & \textbf{F15} & \textbf{F14} & \textbf{F15} & \textbf{F14} & \textbf{F15} \\
        \midrule
        $a$ & -0.0143 & 0 & -0.0119 & -0.0074 & -0.0039 & -0.0072 & -0.0023 & -0.006 & -0.0023 \\
        $b$ & 1.8884 & 1 & 1.7665 & 1.4813 & 1.2645 & 1.4321 & 1.1326 & 1.3605 & 1.1322 \\
        $c$ & -1.8454 & 0 & -2.2813 & -1.5059 & -1.7579 & -0.0765 & 0.4873 & -0.2098 & 0.2721 \\
        $R^2$ & 0.974 & 1 & 0.9883 & 0.9721 & 0.9740 & 0.9659 & 0.9705 & 0.9645 & 0.9696 \\
        
        \toprule
        \textbf{Year} & \textbf{2003} & \textbf{2003} & \textbf{2004} & \textbf{2004} & \textbf{2005} & \textbf{2005} & \textbf{2006} & \textbf{2006} & \textbf{2007} \\
        \textbf{Satellite} & \textbf{F14} & \textbf{F15} & \textbf{F15} & \textbf{F16} & \textbf{F15} & \textbf{F16} & \textbf{F15} & \textbf{F16} & \textbf{F15} \\
        \midrule
        $a$ & -0.0064 & -0.0131 & -0.0127 & -0.0093 & -0.0094 & -0.0116 & -0.0087 & -0.006 & -0.0111 \\
        $b$ & 1.3760 & 1.8092 & 1.7658 & 1.5971 & 1.5570 & 1.7041 & 1.5121 & 1.3510 & 1.6814 \\
        $c$ & 0.5644 & -0.6368 & -0.0817 & -1.8401 & 0.9574 & -0.2285 & 1.7289 & 1.3256 & -0.3691 \\
        $R^2$ & 0.9635 & 0.9602 & 0.9565 & 0.9467 & 0.9531 & 0.9489 & 0.9402 & 0.9229 & 0.9432 \\
        
        \toprule
        \textbf{Year} & \textbf{2007} & \textbf{2008} & \textbf{2009} & \textbf{2010} & \textbf{2011} & \textbf{2012} & \textbf{2013} & \textbf{2013} & \textbf{2014} \\
        \textbf{Satellite} & \textbf{F16} & \textbf{F16} & \textbf{F16} & \textbf{F18} & \textbf{F18} & \textbf{F18} & \textbf{F15} & \textbf{F18} & \textbf{F15} \\
        \midrule
        $a$ & -0.0038 & -0.0039 & -0.003 & 0.0102 & -0.0009 & 0.0030 & -0.0176 & 0.0006 & -0.0194 \\
        $b$ & 1.1971 & 1.1952 & 1.1484 & 0.1829 & 0.9751 & 0.6763 & 1.9709 & 0.8688 & 2.0911 \\
        $c$ & 0.3308 & 0.8991 & 1.2554 & 7.4196 & 1.6559 & 4.6656 & 0.7879 & 2.5010 & -0.3125 \\
        $R^2$ & 0.9334 & 0.9407 & 0.9371 & 0.9261 & 0.9190 & 0.9044 & 0.8723 & 0.9208 & 0.8719 \\
        
        \toprule
        \textbf{Year} & \textbf{2015} & \textbf{2016} & \textbf{2016} & \textbf{2017} & \textbf{2017} & \textbf{2018} & \textbf{2018} & \textbf{2019} & \textbf{2019} \\
        \textbf{Satellite} & \textbf{F15} & \textbf{F15} & \textbf{F16} & \textbf{F15} & \textbf{F16} & \textbf{F15} & \textbf{F16} & \textbf{F15} & \textbf{F16} \\
        \midrule
        $a$ & -0.0209 & -0.0213 & -0.0269 & -0.0218 & -0.0239 & -0.0211 & -0.0199 & -0.0200 & -0.0191 \\
        $b$ & 2.1549 & 2.1562 & 2.4526 & 2.1726 & 2.2963 & 2.1457 & 2.0848 & 2.0940 & 2.0538 \\
        $c$ & 0.7466 & 1.1204 & 1.6481 & 1.0962 & 1.1565 & 1.0301 & 0.7396 & 0.8782 & 0.0231 \\
        $R^2$ & 0.8667 & 0.8531 & 0.8564 & 0.8344 & 0.8369 & 0.8471 & 0.8461 & 0.8516 & 0.8552 \\
        
        \bottomrule
        \end{tabularx}
        \label{fittedParameters} 
    \end{table*}

\subsection{The improved continuity of DMSP-OLS}

    \begin{figure*}[!htbp]
        \centering
        \includegraphics[width=0.55\textwidth]{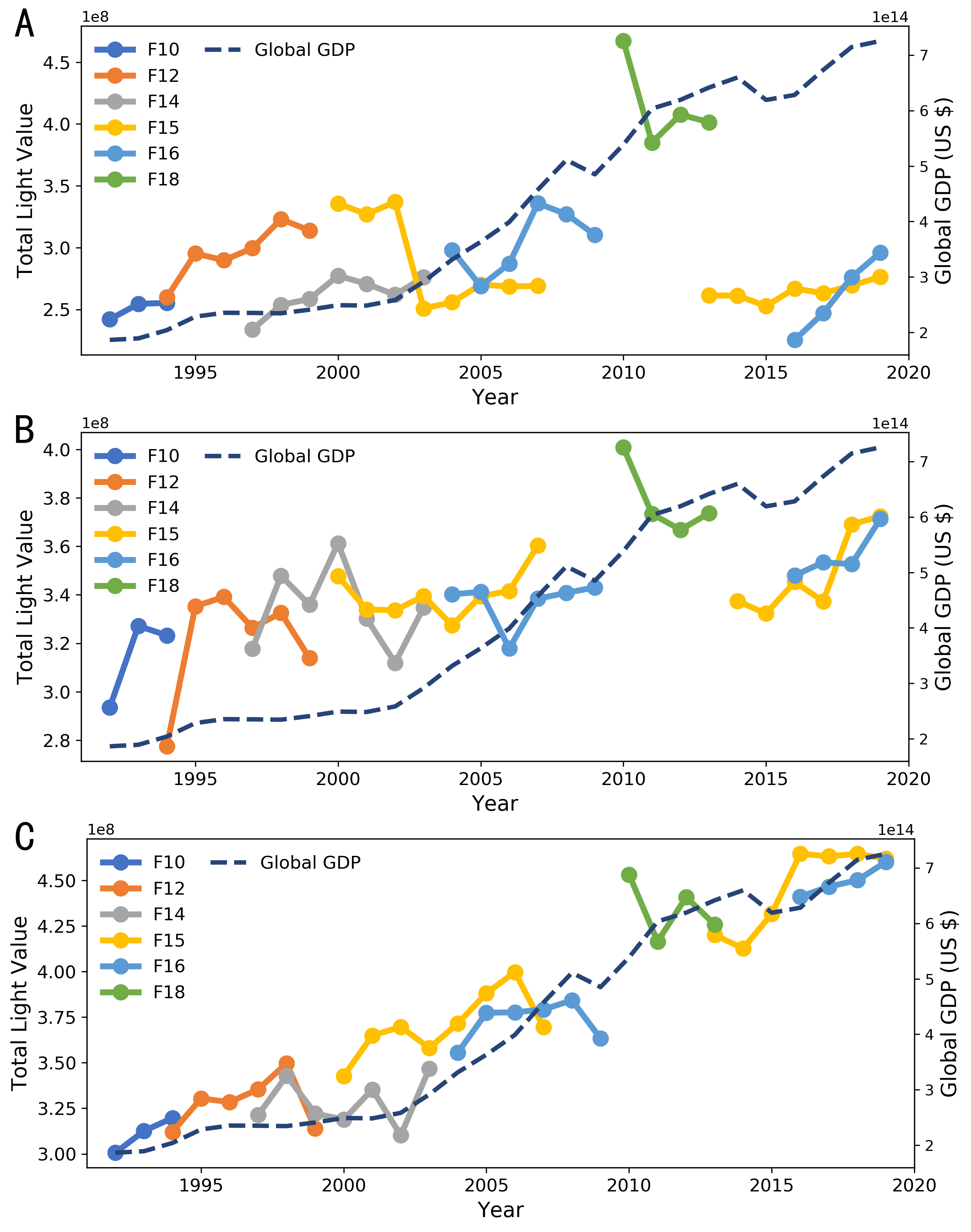}
        \caption{(A) TLV of original DMSP-OLS images. (B) TLV of DMSP-OLS images corrected using Elvidge’s method. (C) TLV of DMSP-OLS images corrected using our method. The dashed curve indicates global GDP.}\label{figA2}
    \end{figure*}

    The global total light values (TLV) in time series can reveal the continuity of DMSP-OLS images. In addition, it has been proven that TLV is well correlated with GDP \citep{elvidge1997relation,sutton2007estimation}. As shown in Figure \ref{figA2} (A), the global GDP has been continuously increasing during the considered period, while the TLVs of original DMSP-OLS images fluctuate significantly and lack continuity. Figure \ref{figA2} (B) indicates the TLVs of the images calibrated using Elvidge’s method, which used the Sicily Island as the calibration field \citep{en20300595}.  The TLVs between 1995 and 2005 are noticeably overestimated. This leads to a wide gap between the TLV and global GDP curves. Moreover, the TLVs after 2014 are underestimated, and hence, there is another wide gap between the TLV and global GDP curves. Figure \ref{figA2} (C) presents the TLVs of images calibrated using our method.  The TLV increases continuously with the increase in global GDP, and there is no significant underestimation or overestimation.  The good correlation between TLV and global GDP indicates that our method can promote the continuity of DMSP-OLS images significantly, and it is better than that of Elvidge’s method.

\newpage
\section{} \label{AppendixB} 
    \subsection{ResNet Architecture}
    \begin{figure*}[!htbp]
    \centering
        \includegraphics[width=1.0\textwidth]{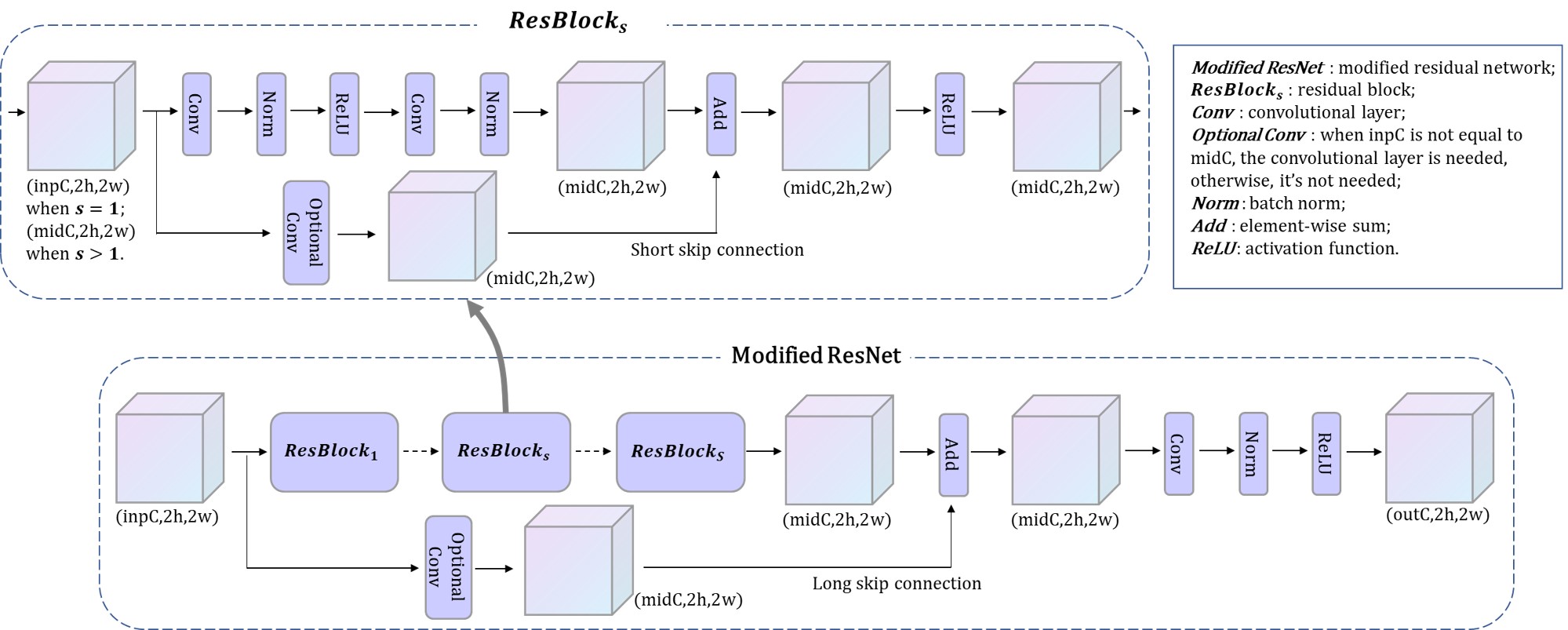}
    \caption{Architecture of the modified ResNet. The modification involves the addition of a long skip connection to the original ResNet.}\label{figB1}
    \end{figure*}

    The modified ResNet is shown in Fig.\ref{figB1}. The sizes of its input, middle, and output tensors are $(inpC,2h,2w)$, $(midC,2h,2w)$, and $(outC,2h,2w)$, respectively.  $inpC$, $midC$ and $outC$ are different in $H^\ast$, $F_3$ and $G^\ast$.  The basic structures of the modified ResNet are some continuously stacked residual blocks ($\emph{ResBlock}$), as presented in the bottom of Fig.\ref{figB1}.  Our modification lies in a long skip connection which sums the input of the first residual block and the output of the last residual block. The long skip connection facilitates a deeper model. If $\emph{inpC}$ is not equal to $\emph{midC}$, an additional convolutional layer ($\emph{Optional\,Conv}$) is needed in the long skip connection to change the tensor’s channel number; otherwise, $\emph{Optional\,Conv}$ is not needed. In the tail of the ResNet, a convolutional layer ($\emph{Conv}$), a batch norm layer ($\emph{Norm}$), and an activation function ($\emph{ReLU}$) are used to produce the output.
    
    A certain $\emph{ResBlock}_s$ is shown in the top of Fig.\ref{figB1}. It consists of two groups of $\emph{Conv}$, $\emph{Norm}$, and $\emph{ReLU}$. The key point is a short skip connection which sums the input of the first $\emph{Conv}$ and the output of the second $\emph{Norm}$. This short skip connection enables the neural network to learn easily.  Similarly, if $\emph{inpC}$ and $\emph{midC}$ are not equal in $\emph{ResBlock}_1$, an $\emph{Optinal\,Conv}$ is also needed in the short skip connection.

\subsection{RCAN Architecture}
    \begin{figure*}[!htbp]
    \centering
        \includegraphics[width=1.0\textwidth]{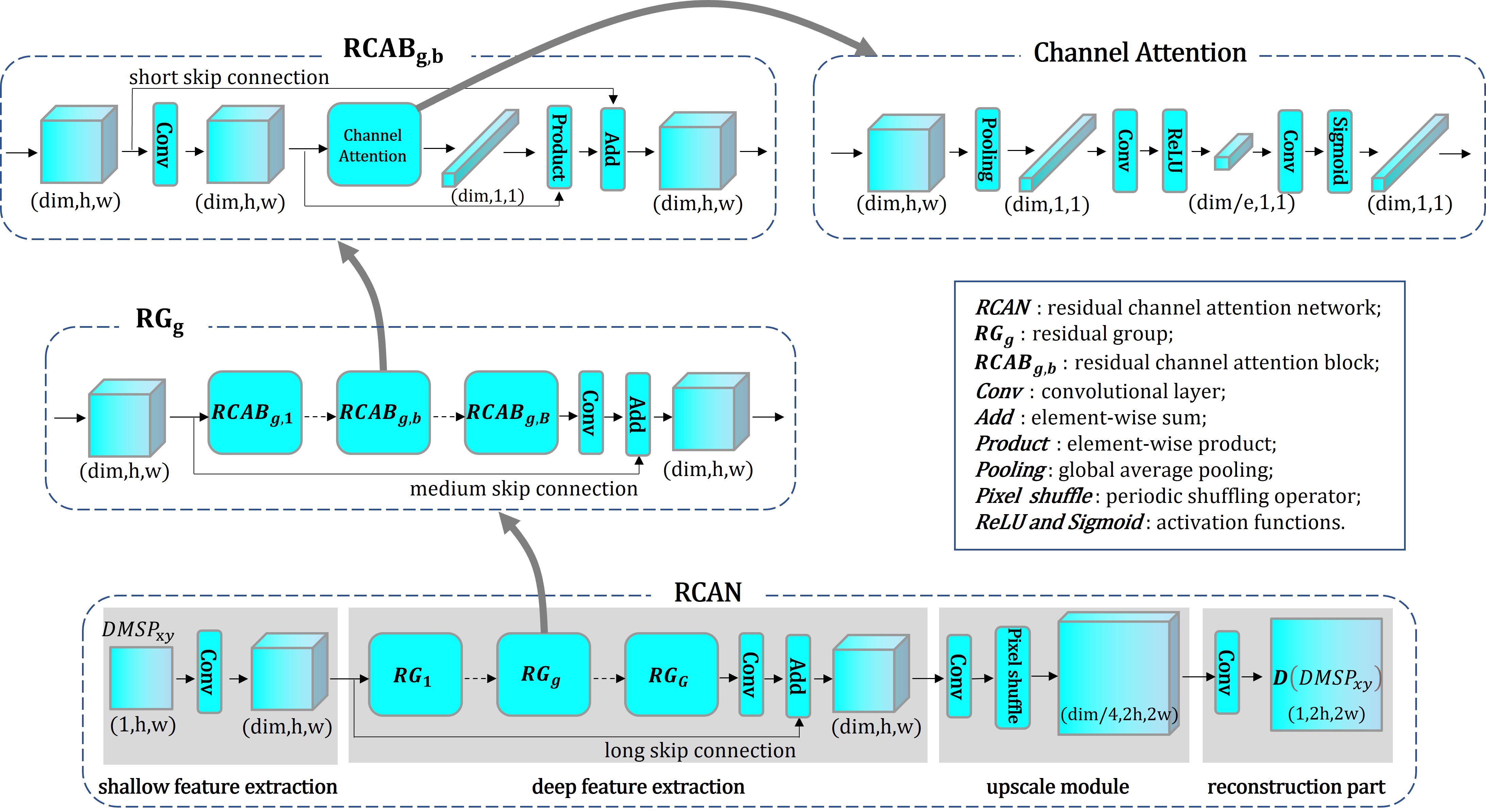}
        \caption{Architecture of RCAN.}\label{figB2}
    \end{figure*}

    The size of the input and output tensors of RCAN are $(1, h, w)$ and $(1,2h,2w)$, respectively. As shown in the bottom of Fig.\ref{figB2}, it can be divided into four parts, including shallow feature extraction, deep feature extraction, upscale module, and reconstruction part.  The shallow feature extraction increases the channel number of the input tensor from 1 to $dim$ by the $Conv$ operation.

    The main part of the RCAN is the deep feature extraction. It further extracts features by stacking $G$ residual groups (RG). To deepen the model, a long skip connection, which sums the input of $RG_1$ and the output of the $Conv$ immediately after $RG_G$, is required. A certain $RG_g$ is shown in the middle of Fig.\ref{figB2}. Essentially, it consists of $B$ residual channel attention blocks (RCAB). A medium skip connection sums the input of the $\emph{RCAB}_{g,1}$ and the output of the $Conv$ immediately after $\emph{RCAB}_{g,B}$. A certain $\emph{RCAB}_{g,b}$ is presented in the upper left of Fig.\ref{figB2}. It first uses a $Conv$ to extract features, then uses a $\emph{Channel\,Attention}$ to weight each channel of the features. Subsequently, a short skip connection sums the weighted features and the input of $\emph{RCAB}_{g,b}$. The $\emph{Channel\,Attention}$ is shown in the upper right part of the figure. In this block, the height and width of the input tensor shrink to 1 through a global average pooling ($\emph{Pooling}$). Then, the channel number shrinks to $\emph{dim/e}$ through a $\emph{Conv}$ and a $\emph{ReLU}$. After that, a pair of $\emph{Conv}$ and $\emph{Sigmiod}$ restores the channel number to $\emph{dim}$, and makes the value of each element between 0 and 1. Thus, the weight of each channel can be obtained.

    The upscale module occurs after the deep feature extraction, as shown in the bottom of Fig.\ref{figB2}. It consists of a $Conv$ and a periodic shuffling operator ($\emph{Pixel\,Shuffle}$). The $\emph{Pixel\,Shuffle}$ reorders the tensor elements to double the height and width of its input tensor. The last part of the RCAN is reconstruction, in which the tensor’s channel number shrinks to 1 by a $Conv$. Finally, the output of the RCAN with a size of $(1,2h,2w)$ is obtained.

\subsection{Hyperparameters}

    $h$ and $w$ were set as 128. For the module $F_3$, $inpC$, $midC$, $outC$, and $S$ were set as 1, 32, 32, and 16 respectively. For the module $H^\ast$, these four hyperparameters were 2, 32, 32, and 32 respectively. For the module $G^\ast$, they were 64, 64, 1, and 32 respectively. In the module $F_1$, $dim$ is 64 and $G$ and $B$ were both 6. In addition, the reduction ratio $e$ in the channel attention of RCAN was set as 16.

    $\emph{Kernel Size}$, $\emph{Padding}$ and $\emph{Stride}$ are three key hyperparameters in a convolution operation.  In the convolutional layers of channel attention, they were set as 1, 1, and 0, respectively. For other convolutional layers in the model, $\emph{Kernel\,Size}$ was set as 3, and the zero-padding strategy was used to maintain the heights and widths of tensors.

\section{} \label{AppendixC} 
    \begin{figure*}[!htbp]
    \centering
    \includegraphics[width=0.79\textwidth]{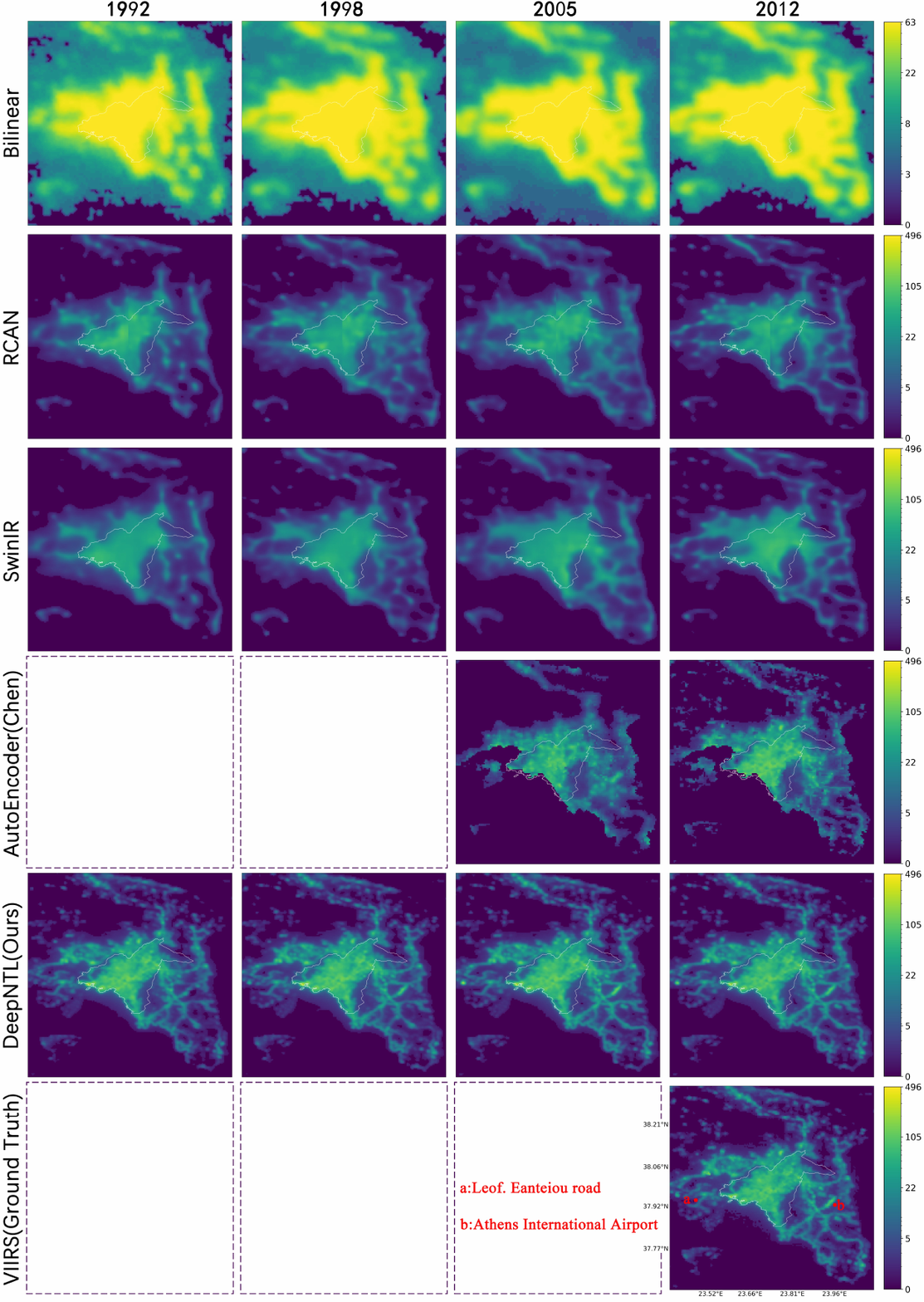}
    \caption{Reconstructed and GT images of Athens. (For better perception, readers are recommended to zoom in on the web version of this figure.)}\label{figC1}
\end{figure*}

\begin{figure*}[!htbp]
    \centering
    \includegraphics[width=0.79\textwidth]{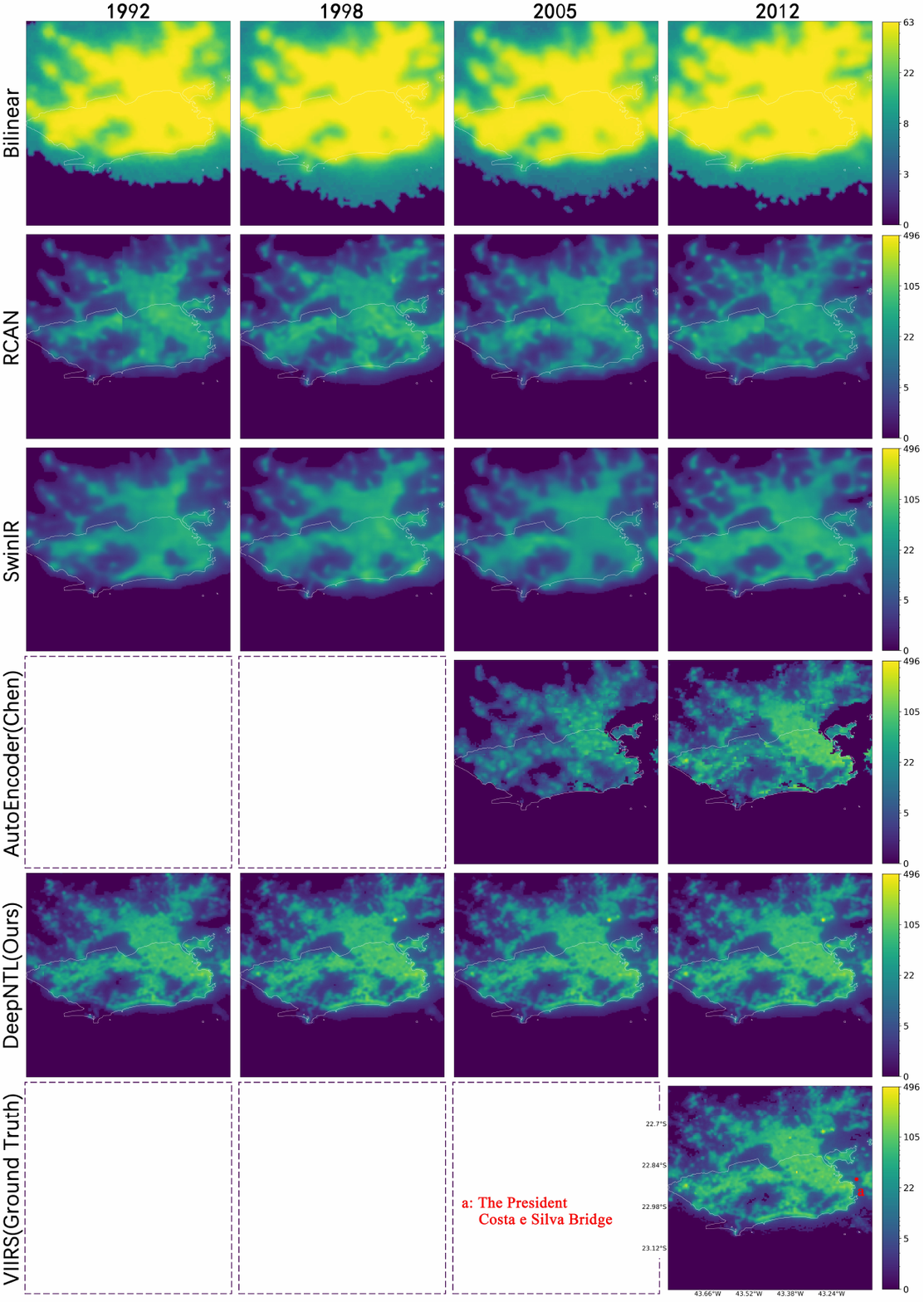}
    \caption{Reconstructed and GT images of Rio de Janeiro. (For better perception, readers are recommended to zoom in on the web version of this figure.)}\label{figC2}
\end{figure*}

\begin{figure*}[!htbp]
    \centering
    \includegraphics[width=0.79\textwidth]{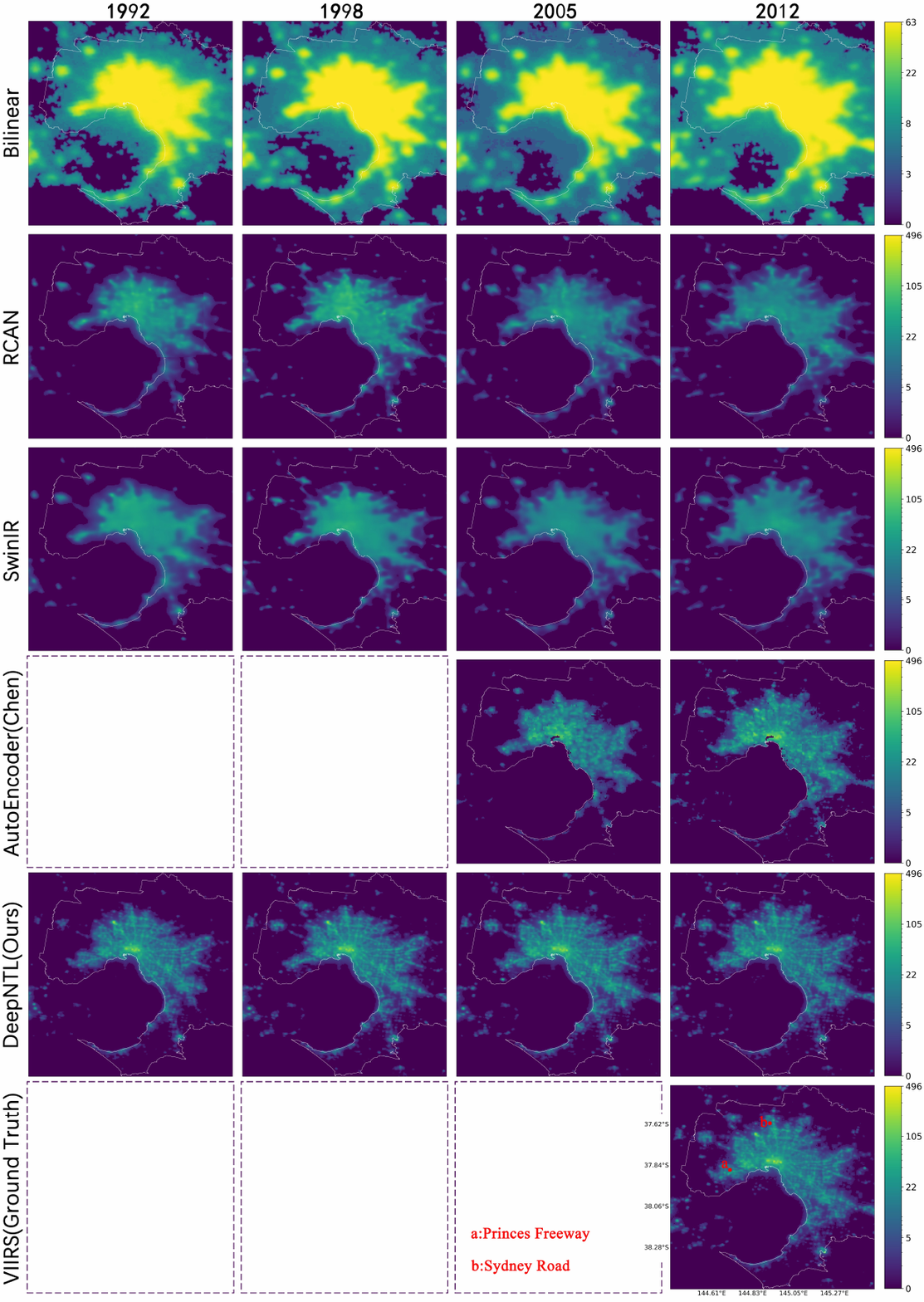}
    \caption{Reconstructed and GT images of Melbourne. (For better perception, readers are recommended to zoom in on the web version of this figure.)}\label{figC3}
\end{figure*}

\begin{figure*}[!htbp]
    \centering
    \includegraphics[width=0.79\textwidth]{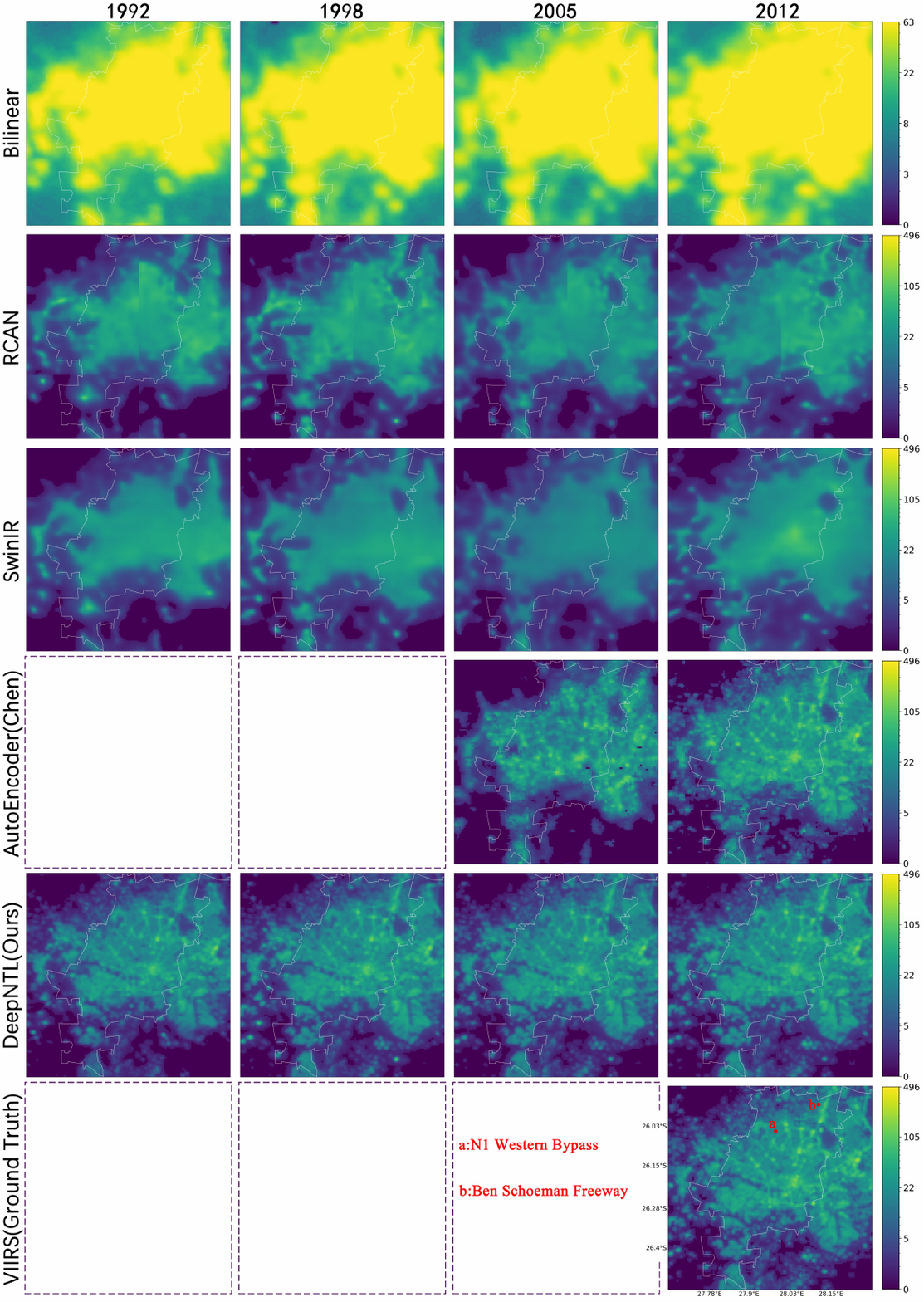}
    \caption{Reconstructed and GT images of Johannesburg. (For better perception, readers are recommended to zoom in on the web version of this figure.)}\label{figC4}
\end{figure*}

As shown in Figure \ref{figC1}, point a in the GT image of Athens indicates Leo Eateiou Road, a seaside road with Salamina Bay to the west. Among these models, only DeepNTL shows it clearly. Point b indicates Athens International Airport, whose construction began in 1996. In the DeepNTL image, the lights at this location were dim in 1992 and brightened significantly after 1998. This is consistent with the facts. In Figure \ref{figC2}, point a in the GT image of Rio de Janeiro indicates the President Costa e Silva Bridge, which straddles Guanabara Bay and connects the cities of Rio de Janeiro and Niteroi. The bridge was completed in 1974. Each year's DeepNTL image shows the bridge clearly, while other models produce unclear images.

As shown in Figure \ref{figC3}, point a in the GT image of Melbourne indicates Princes Freeway, and point b indicates Sydney Road. The bilinear, RCAN, and SwinIR models cannot display them. AutoEncoder can show the two roads only in 2012, while it fails to show them in 2005 due to the decline of its generalization ability. DeepNTL is able to clearly display these two roads each year because its generalization ability is stable. In Figure \ref{figC4}, point a in the GT image of Johannesburg indicates the N1 Western Bypass, which opened in 1975. It presents an approximate circular arc shape. Point b of Johannesburg indicates the Ben Schoeman Freeway, which opened in 1968. This freeway connects Johannesburg with Pretoria in the northeast. Similarly, the images of the first three models are too blurry to clearly show these two roads. AutoEncoder can show them in 2012; however, the decline of its generalization ability resulted in a loss of image clarity in 2005. Only DeepNTL can clearly display these two roads during these years.
\noappendix       

\bibliographystyle{copernicus}
\bibliography{reference}

\end{document}